\documentclass[12pt, oneside, letterpaper]{article}
\usepackage[T1]{fontenc}
\usepackage[margin=1in, letterpaper]{geometry}
\usepackage{setspace}
\usepackage[final]{microtype}
\newcommand{\draftfinal}[2]{\ifdefined\draftversion#1\else#2\fi}
\newcommand{\draftonly}[1]{\draftfinal{#1}{}}

\usepackage{adjustbox}
\newcommand{\titleval}{Reinforcement Learning and \\ Consumption-Savings Behavior}
\newcommand{\authorval}{Brandon Kaplowitz}
\newcommand{\email}{bgk258@stern.nyu.edu}
\newcommand{\school}{New York University}

\draftonly{\setlength{\overfullrule}{5pt}}
\RequirePackage[prologue]{xcolor}
\definecolor[named]{ThesisBlue}{cmyk}{1,0.1,0,0.1}
\definecolor[named]{ThesisYellow}{cmyk}{0,0.16,1,0}
\definecolor[named]{ThesisOrange}{cmyk}{0,0.42,1,0.01}
\definecolor[named]{ThesisRed}{cmyk}{0,0.90,0.86,0}
\definecolor[named]{ThesisLightBlue}{cmyk}{0.49,0.01,0,0}
\definecolor[named]{ThesisGreen}{cmyk}{0.20,0,1,0.19}
\definecolor[named]{ThesisPurple}{cmyk}{0.55,1,0,0.15}
\definecolor[named]{ThesisDarkBlue}{cmyk}{1,0.58,0,0.21}

\definecolor{SchoolColor}{rgb}{0.3412, 0.0235, 0.5490} 
\definecolor{chaptergrey}{rgb}{0.2600, 0.0200, 0.4600} 
\definecolor{midgrey}{rgb}{0.4, 0.4, 0.4}

\usepackage{chemformula}
\usepackage{array}
\RequirePackage[labelfont={bf,sf,small,singlespacing},
  textfont={sf,small,singlespacing},
  margin=0pt,
  figurewithin=section,
tablewithin=section]{caption}

\usepackage{fix-cm}
\RequirePackage[raggedright]{titlesec}

\usepackage{setspace}

\usepackage{lipsum}

\usepackage{setspace}
\usepackage{lipsum}

\input{defs.tex}

\usepackage{comment}
\usepackage{hyperref}
\hypersetup{colorlinks,
  linkcolor=ThesisDarkBlue,
  citecolor=ThesisPurple,
  urlcolor=ThesisDarkBlue,
filecolor=ThesisDarkBlue}

\usepackage[american]{babel}
\usepackage[autostyle=true, autopunct=true]{csquotes}
\MakeOuterQuote{"}

\usepackage[
  backend=biber,
  authordate,
  sorting=nyt,
  maxcitenames=2,
  maxbibnames=9,
]{biblatex-chicago} 
\usepackage{lscape}
\usepackage{varioref}
\usepackage{fancyref}
\usepackage[capitalize]{cleveref}

\theoremstyle{plain}
\newtheorem{theorem}{Theorem}[section]

\theoremstyle{definition}

\newtheorem{definition}[theorem]{Definition}
\theoremstyle{plain}
\usepackage{thm-restate}

\theoremstyle{definition}

\theoremstyle{remark}

\title{\titleval}
\author{\authorval\thanks{\school. Email: \href{mailto:\email}{\email}. \\
    \\
I am grateful to my committee, Jaroslav Borovicka, Thomas Sargent, Thomas Philippon, Andrew Caplin, Peter Ganong, Daniel Greenwald, Corina Boar, Jess Benhabib,  Simon Gilchrist, Benjamin Moll, and Gianluca Violante for helpful comments and suggestions. For valuable discussions and ideas, I thank Kevin Guo, Spencer Kwon, Sobhan Mohammadpour, Matthew Fellows, Samuel Sokota, Christian Schroeder De Witt, Jakob Foerster, Gabriele Farina, Chiara Gardenghi, Man Chon Iao, and Yatheesan Selvakumar. I also thank seminar participants at the North American Summer Meeting of the Econometric Society, NYU Macro Student Lunch, Simons Institute Theory of Reinforcement Learning Workshop, AI4ABM Workshop,  NYU Stern Macro Workshop. All errors are my own.}\\
 \small\href{https://www.bkaplowitz.com/publications/rl-consumption-savings}{[Click here for latest version]}}

\date{\today}
\begin{document}
\maketitle
\begin{abstract}
  \singlespacing
  This paper demonstrates how reinforcement learning can explain two puzzling empirical patterns in household consumption behavior during economic downturns. I develop a model where agents use Q-learning with neural network approximation to make consumption-savings decisions under income uncertainty, departing from standard rational expectations assumptions. The model replicates two key findings from recent literature: (1) unemployed households with previously low liquid assets exhibit substantially higher marginal propensities to consume (MPCs) out of stimulus transfers compared to high-asset households (0.50 vs 0.34), even when neither group faces borrowing constraints, consistent with \textcite{ganongSpendingJobfindingImpacts2024}; and (2) households with more past unemployment experiences maintain persistently lower consumption levels after controlling for current economic conditions, a ``scarring'' effect documented by \textcite{Malmendier2024}. Unlike existing explanations based on belief updating about income risk or ex-ante heterogeneity, the reinforcement learning mechanism generates both higher MPCs and lower consumption levels simultaneously through value function approximation errors that evolve with experience. Simulation results closely match the empirical estimates, suggesting that adaptive learning through reinforcement learning provides a unifying framework for understanding how past experiences shape current consumption behavior beyond what current economic conditions would predict.
\end{abstract}
\singlespacing
\newpage
\onehalfspacing
\section{Introduction}

\subsection{Motivation and Goals}
The overarching goal of this paper is to understand the implications of sophisticated learning protocols for consumption behavior. Specifically, in this project, I focus on two empirical patterns documented in recent years and show that embedding reinforcement learning, a widely used protocol in computer science, in an otherwise standard model of consumption choice can generate these patterns.

In my model, a consumer is represented as an economically rational decision-maker with informational and computational constraints.\footnote{I refer to this consumer as an \emph{agent} throughout.} In particular, the model assumes that the consumer does not know their income process perfectly, nor are they able to solve a Bellman equation to obtain their true value function. Instead, they learn their continuation value over time using reinforcement learning. They form a parametric estimator of their continuation value and update the parameters of this estimator over time based on their experiences.

One crucial assumption that brings considerable tractability is that an agent uses a Markovian formulation of their decision problem, where continuation values depend only on current assets and income.\footnote{Other important assumptions made that are standard in an economics context, but may want to be questioned in this context, include: economic rationality of households (that is, that they are subjective expected utility maximizers); agents forming forward-looking expectations; no special memory, context, replay buffer, attention, or model of the environment; choice of learning algorithm among reinforcement learning algorithms; preferences admitting a unique utility representation and this utility representation being available to households; time-separability of preferences; no cost for computation time or other forms of complexity or access to multiple alternative solution schemes; and no preferences regarding the timing of the resolution of uncertainty, such as ambiguity aversion.} This is restrictive because it presupposes knowledge of the relevant states, but serves as a natural starting point; with this assumption, the problem becomes defined in terms of a Markov Decision Problem (MDP), where reinforcement learning is an appropriate solution procedure.\footnote{%
Relaxing the agent ``magically'' knowing their true state variables represents a natural follow-up. See \cref{app:relaxing-mdp} for further details.}
Two recent empirical findings motivate this analysis. First, \textcite{ganongSpendingJobfindingImpacts2024} study household spending during the COVID-19 recession and document a puzzling pattern: unemployed households with previously low liquid assets exhibited much higher marginal propensities to consume (MPCs) out of stimulus transfers (0.53) than those with previously high assets (0.29), even when the transfers were large enough that neither group faced current borrowing constraints.

Second, \textcite{Malmendier2024} examine how past economic experiences affect current behavior using panel data from the Panel Study of Income Dynamics (PSID) and NielsenIQ Homescan Consumer Panel. They find that households with more past unemployment experiences---either personal or through exposure to high regional unemployment---maintain persistently higher savings rates even after controlling for all current economic conditions including income, wealth, and employment status, and suggest it has a highly persistent effect that they capture via a weighted index. This ``scarring'' effect suggests that past unemployment experiences create lasting behavioral changes beyond their immediate economic impact.

These findings present challenges for standard full-information rational expectation models of consumption, even augmented with borrowing constraints. In particular, past unemployment is associated with both \emph{lower average consumption} and \emph{higher marginal propensities to consume while unemployed}. Reconciling these two patterns jointly is difficult:
\begin{enumerate}[nosep]
  \item Borrowing constraints can generate high MPCs, but agents in \textcite{ganongSpendingJobfindingImpacts2024} exhibit significant MPC heterogeneity (0.53 vs.\ 0.29) despite being \emph{away} from the borrowing constraint.

  \item Higher behavioral heterogeneity or present-bias could generate high MPCs, but would also generate \emph{higher} average consumption, contradicting the scarring pattern.

  \item Standard belief-based learning (anticipated utility, as in \textcite{Malmendier2024}) leads to higher subjective unemployment probabilities, which generates lower consumption but mechanically also generates \emph{lower} MPCs, as I show in \cref{sec:rl-vs-pessimism}.
\end{enumerate}

The model assumes agents approximate their continuation value with a flexible parametric model with a neural network estimate at the rational solution and then proceed to update the parameters governing this model, the weights and biases of the neural network, over the course of their lifetime via gradient descent. To obtain their optimal consumption choice, agents fit a polynomial to the output of the neural network. For initialization, the model assumes each agent begins their life with an estimate of the continuation value that coincides with its rational expectations counterpart. This means that deviations from rational expectations in the model arise from experiences and learning dynamics rather than initial conditions. A key parameter is the learning rate, which controls the rate at which agents update the parameters in response to new observations. The baseline parameterization uses the slowly decaying learning rate \(\lambda_t=\lambda_0/\sqrt{t+1}\). This is a standard slowly decaying step-size schedule. The exact Robbins-Monro square-summability condition \parencite{Robbins1951} requires a slightly faster decay, \(\lambda_t=\lambda_0(t+1)^{-\rho}\) with \(\rho>1/2\) to guarantee convergence, but in long-run simulations the baseline schedule appears to converge back toward the rational solution.  
I simulate the model with standard quarterly consumption-savings parameters heavily drawn from \textcites{ganongSpendingJobfindingImpacts2024, Krueger2016}. I examine their consumption profiles over 50 quarters.

I use the calibrated model to replicate two results that connect to recently documented empirical facts.
First, consistent with \textcite{ganongSpendingJobfindingImpacts2024}, agents who previously held low liquid assets exhibit substantially higher marginal propensities to consume (MPCs) out of stimulus transfers compared to previously high-asset agents, even when transfers of cash are large enough that neither group is currently borrowing constrained.
Using data generated via simulation, I find estimates very similar to those found by \textcite{ganongSpendingJobfindingImpacts2024}. Second, following \textcite{Malmendier2024}, agents with more past unemployment experiences maintain higher savings rates, significant even after controlling for current assets and income and dropping the two most recent observations---a \emph{scarring} effect from past unemployment experiences.

Thus, this paper presents RL as an alternative unifying explanation for these facts. It does not rely on ex-ante heterogeneity---the hypothesis in \textcite{ganongSpendingJobfindingImpacts2024}---which cannot account for the experience effects observed in \textcite{Malmendier2024}. RL is also distinct from the scarring mechanism in \textcite{Malmendier2024}, as scarring focuses solely on changes in future expected unemployment probabilities. Empirically, my mechanism generates both lower consumption levels and higher MPCs, consistent with the data. In contrast, increased subjective unemployment risk (as learned through scarring) causes MPCs and consumption levels to fall together. This pattern is inconsistent with the higher MPCs that \textcite{ganongSpendingJobfindingImpacts2024} find among agents with lower endogenous prior assets---consistent with agents who have experienced past unemployment and depleted their savings through consumption-smoothing.

\section{Literature}\label{sec:literature}
This paper is related to several strands of the literature. The substantive question bears a direct connection to a large, long-standing body of work studying the effects of incomplete information and learning on consumption behavior.

\paragraph{Learning and Consumption} One influential strand explores the implications of unknown income transition probabilities, where an agent learns over time about their income process but perfectly solves their dynamic optimization problem.\footnote{By iterating forward on the perceived income transition probabilities to a fixed point.} This approach includes \textcite{Cogley2008}, who introduced \emph{anticipated utility} models to macroeconomics, where agents use a Bayesian approach to learn about an underlying exogenous stochastic process for income or some other state they have to forecast, but then solve their value function to a fixed point, assuming no further learning happens from tomorrow onward. While specifics of the learning process and economic settings differ, \textcites{Friedman1957, Muth1960, Jovanovic1979, Marcet1989, Evans1994, Guvenen2007, Kozlowski2020} all fall into this category. \textcite{Malmendier2024}, who document the scarring fact mentioned earlier, use a similar model to explain their fact.

This paper differs from that work in that it does not assume that agents learn about their income process or solve a fixed-point problem every period. Instead, they directly estimate the value function associated with different choices each period. In particular, this paper never requires them to solve their value function forward in time via iteration, given a particular set of beliefs. However, agents remain forward-looking in that they always are forming estimates of their expected value function and making decisions based on estimated future expected utility realizations.

\paragraph{Reinforcement Learning} The second large strand of literature is the reinforcement learning literature, which has been widely applied in computer science and to some extent in economics. This literature focuses on estimating the value function of a Markov decision process (MDP) and using this to make decisions about how to act in the future. \textcite{suttonReinforcementLearningIntroduction2018} provide a comprehensive overview and introduction to this literature. Additionally, there has been a large body of recent literature that suggests that humans learn in ways that resemble (deep) reinforcement learning. \textcites{Daw2011, Dabney2020, Botvinick2020, Muller2024, Kasdin2025, Masset2025, Sousa2025}

In economics, the most relevant papers focusing specifically on reinforcement learning as a learning mechanism are the works by \textcites{Barberis2023, Ilut2024}. \textcite{Barberis2023} study how some puzzles in finance may be explained by a combination of ``model-free'' learning (represented by a variant of \(Q\)-Learning) and ``model-based'' learning (represented by the correctly specified \(Q\)-function, given a learned perceived distribution over returns, akin to adaptive learning). \textcite{Ilut2024} study a particular Gaussian process-based variant of \(Q\)-Learning as a theory for decision-making and explore how agents may trade off between collecting samples and using a linear model to update their value functions over time. They interpret their model via a unified Bayesian lens that determines how optimally agents should decide when to use the model-free versus model-based approach. They also look at how additional factors, such as rational inattention, influence how agents trade off between the model-based and model-free approach.

This paper is motivated by two empirical facts from the recent literature, which I will briefly summarize here. The first fact is from \textcite{ganongSpendingJobfindingImpacts2024}, who study how households' spending patterns during the COVID-19 recession varied based on past financial wealth, and the second fact is from \textcite{Malmendier2024}, who study how past unemployment experiences affect current consumption behavior.

\paragraph{Marginal Propensities to Consume} During the COVID-19 recession, between March 29, 2020 and July 31, 2020, Congress temporarily added \$600 Federal Pandemic Unemployment Compensation (FPUC) top-ups to regular UI with the CARES Act.

The paper leverages a unique natural experiment created by administrative processing delays in state unemployment insurance systems. This quasi-random variation caused otherwise-similar unemployed workers to receive benefits at different times in different states. \Citeauthor{ganongSpendingJobfindingImpacts2024} use this to identify causal effects of benefit receipt on household spending patterns and to explore how these effects differ based on household liquid asset levels, particularly looking at differences in marginal propensities to consume (MPCs) across households with different histories. The paper examines differences in consumption due to staggered stimulus receipt across matched otherwise similar households in different states and defines MPCs as the additional share consumed by households that received stimulus payments within a month of stimulus receipt relative to those that didn't, as a share of the total stimulus payment received. This is meant to estimate the slope of the consumption function---what fraction of a dollar of additional income is immediately consumed by households.
This provides the first fact to target.
\begin{tcolorbox}
  \textbf{Fact \#1:} During COVID-19, the average MPC out of stimulus transfers for previously low-asset unemployed households is substantially higher (0.53) than that of previously high-asset unemployed households (0.29), in \citeauthor{ganongSpendingJobfindingImpacts2024}'s preferred specification.
  This result holds even though, at present, neither type of agent is borrowing constrained \parencite{ganongSpendingJobfindingImpacts2024}.
\end{tcolorbox}
This is surprising because much previous work argued that high MPCs
observed during unemployment resulted from agents being borrowing-constrained and lacking sufficient liquid assets to smooth consumption.

In fact, benefits were large enough that unemployed households temporarily moved up the asset distribution relative to before they were unemployed, with the median household moving from the 38th percentile to the 63rd percentile of the liquidity distribution, an adjusted measure of bank account balance sheets. Because MPCs fall as asset levels increase in a full-information rational expectations model with borrowing constraints, households with significant asset buffers should not exhibit large differences in MPCs even if there remain persistent differences in their asset levels, raising a puzzle of what causes this difference. Additionally, despite high liquidity, these households also exhibited high marginal propensities to consume relative to past estimates from the literature such as \textcite{Kaplan2022}.\footnote{%
Borrowing constraints alone result in lower MPCs in calibrated simulation, between 5\% and 30\% for quarterly MPCs out of an unexpected windfall of \$500, with the lower end being estimates for average quarterly MPCs in models without illiquid assets, 25\% as the MPC for the poor hand-to-mouth household, and 30\% for the wealthy hand-to-mouth household in two-asset variants with a liquid and illiquid asset.\ According to \textcite{Kaplan2022}, in turn citing \textcites{Jappelli2010, Havranek2020}, a large body of literature argues in favor of MPCs out of transitory income changes of \$500--\$1,000 of 15--25\%.\ \parencite{Kaplan2022}.}

Furthermore, if agents are rational and fully informed, previous asset conditions from several years ago should have little effect on today's decision-making since assets fully capture the impact of the history on states today.

\paragraph{Scarring}\textcite{Malmendier2024} investigate whether personal experiences of economic downturns create persistent behavioral changes that extend beyond the immediate economic impact. Using panel data from the PSID (1993--2013) and NielsenIQ Homescan Consumer Panel (2004--2013), they construct an experience-based index that captures households' exposure to unemployment over their lifetime from both personal and macroeconomic sources, such as regional or national unemployment rates, with more recent experiences receiving greater weight through a weighting scheme that places less weight the further ago an unemployment experience occurred, using a weighted averaging scheme. The core question of the paper examines whether past unemployment experiences affect current consumption behavior even after controlling for all observable economic characteristics, including current income, wealth, employment status, and demographic factors. That is, among two otherwise identical households \emph{today}, does one household's behavior change when they had unemployment experiences in their past or lived through periods of higher national unemployment? By regressing household consumption on this unemployment experience index while holding constant current economic conditions, they can identify whether past hardships leave lasting behavioral ``scars'' that manifest as persistently lower consumption and higher savings.
This provides the second fact to target.
\begin{tcolorbox}
  \textbf{Fact \#2:} Comparing households by their previous unemployment experiences, using a weighted moving average index of unemployment that decays in how long ago the unemployment experiences occurred, the regression coefficient for consumption on past unemployment experiences is negative after controlling for all observables (including current assets of the household) \parencite{Malmendier2024}.
\end{tcolorbox}
These two facts might appear to contradict each other, but they do not for several reasons.
One, I will especially emphasize is the difference in focus on marginals versus averages for the consumption function---that is the difference between the slope of the consumption function and its level.

\paragraph{Marginals versus Averages} \textcite{ganongSpendingJobfindingImpacts2024} is about differences in changes in consumption between two groups (the change in consumption before and after a
shock occurs).
In contrast, \textcite{Malmendier2024} is about differences in average savings levels
(without differences over time).
The first deals with the slope of a line that connects consumption policy points, whereas the other deals with the policy level itself, conditioned on different historical experiences.\footnote{%
  Other differences across the two papers include:

  \textbf{Timescale}:
  The timescale for each claim is very different.
  The first claim spans from a few months to two years.
  The second covers several years to as much as forty years.

  \textbf{Asset Controls}:
  The first claim does not control current assets, whereas the
  second aims to do so.
  Both approaches have implications that might introduce possible
  confounders or selection effects. For example, in the first model, high state-dependent consumption
  could be concentrated among certain households, leading to the
  low asset holdings observed. However, as emphasized already and noted by \textcite{ganongSpendingJobfindingImpacts2024}, due to the focus on marginals, this would not be sufficient to explain the gap of 0.24 in MPCs, since both households should have minimal differences in MPCs being far away from the borrowing constraint under a full-information rational expectations model without behavioral effects.

  \textbf{Group Definitions}:
  The first claim concerns households that previously held low
  liquid assets versus those that held high liquid assets.
  The second pertains to previous unemployment experiences.
  Although correlated, these two definitions are not identical.
  So far, my analysis has treated both claims as referring to past unemployment.
  Other explanations might introduce further degrees of freedom (e.g., intergenerational wealth), but these may be difficult to justify or may complicate the model substantially.
}

I propose a mechanism, reinforcement learning, that is consistent
with both facts. It will generate a decrease in average consumption levels and an increase in MPCs (a shift downward in the consumption policy and a steepening) after periods of unemployment.

The project aims to explain each fact based on a single
learning-based mechanism. This approach ensures that agents do not require ex-ante heterogeneity to explain each result. Instead, experiences are sufficient to generate the ex-post heterogeneity observed in the data.

\section{Model}

Agents are infinitely-lived and discount future utility at rate \(\beta<1\).
Each agent decides the fraction of their cash-on-hand to consume or save in a risk-free asset with a fixed return \(R\) per period. Each agent is fully rational each period in all variants of the model, exactly making the optimal choice, given current estimates.

An agent's cash-on-hand comprises income from two sources: a stochastic labor income process that evolves independently over time and is independent of the agent's choices and a deterministic and perfectly controllable financial income deriving from the agent's asset holdings this period.

\paragraph{Notation}
Denote cash-on-hand \(x\), share of cash-on-hand consumed \(\psi \),
savings choices \(a'\), labor income \(y\), risk-free assets \(a\),
and fixed return on assets per period \(R>1\).
Throughout the paper, prime notation (\('\)) will denote next-period values. Current-period variables appear without primes, next-period with primes. Time subscripts are avoided except where explicitly tracking multiple periods is necessary.
I denote savings as \(a'\) because savings chosen today become assets tomorrow.

\paragraph{Definitions and Relationships}
Cash-on-hand is defined as \(x \coloneqq y + Ra\), consumption is the
share of cash-on-hand consumed \(c=\psi x\), and savings is the
residual, \(a'=(1-\psi)x\).\footnote{Note that in this definition, I
  implicitly assume that agents want to consume as much as possible; mathematically, utility is strictly monotone in consumption and
economically, agents' preferences satisfy local non-satiation.}
Lastly, I note that \(a'_t(a_t,y_t)=a_{t+1} \), that is the agent's \(a'\) choice today, given a realization of assets and income, becomes the agent's realized assets for tomorrow. Savings choices are made before knowing the realization of tomorrow's
income state \(y'\), which, conditional on today's income,
is the realization of a Bernoulli random variable.
\paragraph{Income Process} Income transitions are captured by a
first-order Markov transition matrix \(P\), such that \(y' \sim P_{yy'} \).
Where relevant, I denote the transition probability \(p(y'=y_i|y=y_j)
\) as \(p_{ij} \), \(i,j \in \{e,u\} \), where \(y_e\) is the high-income state (employment) and \(y_u\) is the low-income state (unemployment).
The model assumes \(\beta R <1\) so that the rational information expectations
agent has no motive to accumulate assets beyond smoothing consumption
in response to risky shocks and ensuring sufficient assets to avoid
hitting their borrowing constraint in the future (precautionary savings).
I explicitly consider two types of agents: an agent referred to as the \emph{rational benchmark} with complete and perfect information about their decision problem and that solves their decision problem optimally, and a learning agent, who I refer to as the \emph{reinforcement learner}, that lacks this information, instead using reinforcement learning to estimate their expected-value function each period.
\subsection{Decision Problems}
\paragraph{Rational Benchmark}
In the rational benchmark, the agent's consumption choice can be represented via the Bellman equation.
\begin{equation}
  \begin{aligned}
    V(a,y)
    =
    \max_{c} u(c)
    & +\beta \, \mathbb{E}_{y'|y}
    \bigl[V(a',y')\bigr],          \\
    & a' \ge \underline{a}        \\
    & c = Ra + y - a'             \\
    & c> 0                        \\
    & y' \sim P_{yy'},
  \end{aligned}
\end{equation}
where \(u(c) \) is the agent's utility function for a strictly positive
consumption level \(c\), \(V(a,y) \) is the value function at the agent's state variables, the
asset, income pair \(a,y\), and
\(\mathbb{E}_{y'|y} \) is the expectation over the distribution of the income tomorrow given today's income state. I set \(\underbar{a}=0\) to capture the agent's no-borrowing constraint.

\paragraph{Simulated Agent} Under reinforcement learning, the model assumes the agent cannot exactly solve for their value function via iteration to a fixed point. This could be due to some combination of the fact that the agent does not know their
income process \(P_{yy'} \), or even the family of distributions
their income process comes from, or cannot solve this fixed-point problem due to computational complexity. However, they can still evaluate their utility
function \(u(c) \). This is intended to reflect a real-world setting
where (1) a household has considerable incomplete information about
their income process, and (2) the household has not been able to get
income data from others.

Because \(V\) is defined as the fixed point of the Bellman
operator---which depends on knowing \(P_{yy'}\) to evaluate the right
side of, which the agent is unable to evaluate exactly, and
\(c^{*}(a,y)\) as the optimal consumption policy that
satisfies this functional equation---the agent can no longer solve exactly
for \(V\). The model assumes that agents do not even know the form of
\(P_{yy'}\), which rules out anticipated utility-based learning,
correctly specified regression, etc. This is a strong assumption in that it assumes agents have no knowledge about the underlying income process (rather than just misspecified knowledge or some form of sampling-based estimation) and that agents cannot solve this fixed problem. I further assume that agents learn solely from their own observations, ruling out the possibility of learning from other agents.

Agents have a perceived expected value function (EV), which I will denote as
\(\widehat{EV}(y,a'; \phi)\).
The parameter \(\phi\) denotes a set of parameters associated with a functional approximator. For example, in the case of a neural
network, which I will use, these will represent the hidden weights
of each layer, while for a polynomial these represent the
coefficients on each term.


Given this perceived EV, I can define \(Q(\cdot)\) associated
with any \(a'\) as follows
\begin{align}
  Q(a,y,a'; \phi)
  =
  u(Ra + y - a')
  +
  \beta \, \widehat{EV}(y,a'; \phi)
\end{align}
\(Q \) gives the expected value of taking arbitrary \emph{feasible} savings choice \(a'\) 
given the estimate \(\widehat{EV}(\cdot)\). The agent then forms their consumption policy as
\begin{align}
  \label{eq:policy}
  {a'}^{*}(a,y; \phi) &\coloneqq \argmax_{a'}Q(a,y,a';\phi) \\
  c^{*}(a,y;\phi) &\coloneqq Ra+y - {a'}^{*}(a,y; \phi)
\end{align}
The agent learns by updating its conditional expected value
\(\mathbb{E}_{y'|y}[V]\) and using this, in turn, to update their \(Q\).

\subsection{Learning}

This section describes how actions unfold and learning occurs in each period. The agent enters the current period with an asset level \(a\) and an income level \(y\). For expositional clarity, I subdivide each period into two stages, day and night.
During the day, the agent chooses \(a'\).
During the night, the agent learns the realization of \(y'\). At that point, he also observes flow utility \(u\) and computes their updated \(\widehat{EV} \), which in turn updates \(Q \) and the agent's policy.
After updating their policy, at the beginning of the following day,
the agent executes the updated policy.

\paragraph{Day} During the day, they choose \cref{eq:policy}.

\paragraph{Night} At night, the agent experiences their flow utility \(u(Ra+y -{a'}^{*}(a,y; \phi))\) and learns their following period's income realization \(y'\). Using this, they form the following \emph{empirical} estimate of the \(Q\)-value.
\begin{align}
  Q^{empirical}(a,y,a',y';\phi)
  \coloneqq
  u(Ra + y - a')
  +
  \beta
  \max_{a''} Q(a', y', a''; \phi).
\end{align}

The temporal difference (TD) error is defined as:
\begin{align}
  \epsilon(a, y, a', y'; \phi)
  & \coloneqq
  Q^{empirical}(a,y,a',y';\phi) - Q(a,y,a'; \phi)
\end{align}
Using the definition of \(Q(a,y,a'; \phi)\) and plugging in for utility, the flow utilities for the present period \(u(Ra+y-a')\) cancel, yielding
\begin{align}
  \epsilon(a,y,a', y' ; \phi) & =
  \beta
  \left[\max_{a''}
    \bigl(
      u(Ra'+y'-a'')
      +
      \beta \, \widehat{EV}(y',a'';\phi)
    \bigr)
    -
  \widehat{EV}(y,a';\phi)\right].
\end{align}
I use this \(\epsilon\) to adjust \(\phi\) using a semi-gradient TD step. Equivalently, this is gradient descent on the squared TD error treating the bootstrap target \(\max_{a''}Q(a',y',a'';\phi)\) as fixed.
The temporal difference error \(\epsilon\) captures the difference between agents' realized value from taking an action and agents' prior expectation of the value of the choice. It therefore represents a kind of ``felicity'' surprise the agent has received from a particular choice.
The parameter update is given by:
\begin{align}
  \phi'
  =
  \phi
  +
  \lambda_t
  \epsilon(a,y,a',y';\phi)
  \nabla_{\phi} \widehat{EV}(y,a';\phi).
  \label{eq:learn}
\end{align}
They then revise the policy as
\begin{align}
  {a'}^{*}(a,y; \phi')
  & =
  \argmax_{a'\in[\underline{a},\overline{a}]}
  Q(a,y,a';\phi'), \\
  c^{*}(a,y; \phi')
  & =
  Ra + y - {a'}^{*}(a,y; \phi').
\end{align}
Here \(\eta_t=\eta_0(t+1)^{-\rho}\) is the primitive step size,

and I absorb the positive factor \(\beta\) from

\(\nabla_\phi Q(a,y,a';\phi)=\beta\nabla_\phi \widehat{EV}(y,a';\phi)\)

into the learning rate:

\[
  \lambda_t \equiv \beta \eta_t =\lambda_0(t+1)^{-\rho}, \qquad \lambda_0=\beta\eta_0>0.
\]
For the Robbins-Monro square-summability condition, \(\rho \in (1/2,1]\), but the baseline simulations use the boundary condition \(\rho=\frac{1}{2}\).
The baseline simulations also use the more learning-rate independent ADAM instead of a plain semi-gradient step.
In simulation, unemployment experiences cause consumption to decrease and savings to increase, while the estimated MPC increases. For the remainder of this paper, I use the following definition for MPC.
\begin{definition}\label{def:MPC}
  The marginal propensity to consume out of a transfer \(\tau \) is
  \begin{equation}
    MPC(a,y; \phi)
    \coloneqq
    \frac{c^{*}(a+\tau,y; \phi)-c^{*}(a,y; \phi)}{\tau}.
  \end{equation}
\end{definition}
\subsection{Parameterization of EV}
The model assumes the agent approximates \(\widehat{EV}\)
with a two-layer neural network with rectified linear unit activations, abbreviated as \(\relu\). I use this due to its desirable properties in approximating complex functions efficiently, because it is the most popular existing activation function \parencite{Rasamoelina2020}, and because it has established guarantees both on convergence to a global optimal fit in the mean-squared error sense in the regression case, and on convergence to an optimal policy in the \(Q\) learning case \parencite{Xu2020}.
The ReLU network is trained in a supervised fashion to imitate the
rational benchmark at time \(0\) to achieve a precise fit to the rational benchmark's continuation value function \(EV(y,a')\).
This specification abstracts from experimentation motives. This is in the spirit of the anticipated utility formulation adopted by much of the learning literature.
Importantly, by using a parameterized estimator of \(\widehat{EV}\), I partially capture the benefits of future exploration on policies to the extent these can be approximated in the \(\widehat{EV}\) estimate. However, these only approximately do so, and the requirement to always use exactly optimal policies rules out the possibility of explicit exploration heuristics, designed to represent the option value from how observations will affect the future evolution of uncertainty given policy choices.
\subsection{Smoothed Neural Network}

A neural network with \(N\) layers can be represented as follows.
Given an input \(X^0\), each layer applies the recursion:
\begin{align}
  X^{i+1}
  =
  \sigma^{i}\bigl(A^{i}X^{i} + b^{i}\bigr),
  \quad
  \forall i \in \{0,\dots,N-1\},
\end{align}
where \(X^0 = {[y,a']}^T\) serves as the argument for
\(\widehat{EV}(y,a') \).
\(A^0\) is an \(h \times 2\) matrix of hidden weights, \(b^0\) is
an \(h \times 1\) vector of hidden biases, and \(X^1\) are the resulting hidden layer latent activations.
Each \(A^i\) for \(i=1,\dots,N-1\) is an \(h \times h\) matrix, while
\(A^N\) is \(1 \times h\), producing the agent's final estimate of
\(\widehat{EV}(y,a') \) for an input pair of \((y,a')\).
The parameter vector \(\phi\) is the concatenation of all these weights and biases as
\begin{align}
  \phi =
  \begin{bmatrix}
    \text{vec}(A^0) \\
    \text{vec}(A^1) \\
    \vdots \\
    \text{vec}(A^{N-1}) \\
    b^0 \\
    b^1 \\
    \vdots \\
    b^{N-1}
  \end{bmatrix}
\end{align}
\(\sigma^i\) is an activation function applied element-wise.
For all layers, I use the ReLU function,
\[
  \relu(x) = \max(x,0).
\]

Because ReLU is non-smooth, the resulting network is also non-smooth,
as are the agent's policies. Non-smooth policies may generate jumps in
consumption. To ensure strict monotonicity in consumption, in line
with the economic literature, the model assumes a polynomial fit to the output of the value function produced by the ReLU network, which the agent takes as their value function values
when considering policies at states other than their
current state in order to generate monotone policy functions.
However, when just computing consumption at their exact current state/realized histories, the agent does not utilize this fit, including when computing realized MPCs out of transfers. This approach retains some of the flexibility of ReLU while smoothing the final output, in particular ensuring strict monotonicity of the consumption policy, which otherwise exhibits local oscillations. In \cref{app:unsmoothed}, I present the results of the neural network fit alone.

\section{Parameterization}

All dollar amounts are normalized to quarterly income units for an employed agent at the time of the COVID-19 stimulus payment. Income is chosen to match \textcite{ganongSpendingJobfindingImpacts2024}'s average ``primary'' income. This represents 0.44 of total household income in their sample and captures the income earned while employed that is not earned while unemployed and still waiting for UI benefits. This is estimated at a nominal basis of \$9,042 on a quarterly basis, which becomes the simulation's baseline unit of 1. Additional
unemployment benefits, on a quarterly basis, were \$7,200 (\$600 a
week). In \textcite{ganongSpendingJobfindingImpacts2024}'s sample,
these come out to 34.5\% of total average income, 78.4\% of primary income. Unemployment is set at the replacement rate of
47.27\% or around \$4,275 on a quarterly basis, again following
\textcite{ganongSpendingJobfindingImpacts2024}. I consider possible
savings choices between 0 and 4.5 times the agent's average quarterly income
(approximately between 0 and \$90,000). I uniformly subdivide this
grid into 8,750 feasible savings choices for policy decisions or
approximately \$10.25 increments of savings, on a quarterly basis.

\subsection{Initialization}
\paragraph{Neural Network}
During training for the initial fit, the ReLU network takes the average MSE over minibatches of size 32 collected from 500 savings choices and two income choices and does stochastic gradient descent on \(\phi\) with ADAM\@, until these attain a tight fit on the rational solution value function globally. I initialized the agents' weights \(A^i\) for the neural network, before the fit from i.i.d.\ draws of a Gaussian with 0 mean and \(\sqrt{2}\) standard deviation, as is required for the theoretical guarantees of \textcite{Li2017} to apply. I initialized the agents' biases \(b^i\) at 0.\footnote{\textcite{Xu2020} also require a Gaussian initialization for their results to go through. They require a slightly different variance to get their theoretical analysis and particular convergence rates. I prefer \textcite{Li2017}, because \textcite{Xu2020} also require the assumptions of the network to be sufficiently overparameterized, which can be difficult to tell if it is satisfied.}

The network was trained to minimize the maximum error over held-out asset values not included in the training sample. I stopped training when the maximum error of the neural network fit fell below \(2.5\times 10^{-3} \),
below which achieving a given error at the current neural network size became noticeably more challenging.

Agents' initial asset holdings are drawn from an empirical distribution calibrated to the 2016 Survey of Consumer Finances (SCF). I use median estimates in the interquartile ranges to get financial asset values at the 12.5th, 37.5th, 62.5th, 87.5th, and 95th
percentiles, which are then normalized to be multiples of quarterly income, the simulation's numeraire. For the lower 87.5 of the distribution, I apply piecewise linear
interpolation between these data points to estimate the inverse CDF\@. For the upper tail, I fit a Pareto distribution with shape
parameter \(\alpha\) estimated from the ratio of the 95th to 87.5th percentile values using \(\alpha =
-\ln(1-0.95)/\ln(Q_{95}/Q_{87.5})\), where \(Q_p\) denotes the asset value at percentile \(p\). This hybrid
approach captures both the empirical distribution of typical households and the heavy-tailed nature of wealth
concentration, with the Pareto tail accounting for extreme wealth holdings above the 87.5th percentile. Unlike the raw SCF, I truncate agents' distribution at 0, as no borrowing is allowed in the simulation, so I force the 0th percentile to be 0. I do not adjust other percentiles to compensate due to lacking the 0th percentile summary statistic. Due to the piecewise linear interpolation, my method may oversample the first octile of the distribution, slightly compared to the population distribution. Sampling occurs via the inverse CDF method, drawing uniformly from the interval [0,1] and applying the inverse linearly-interpolated CDF to obtain asset values. This ensures that the initial asset distribution reflects the empirical distribution of household assets while maintaining the no-borrowing constraint.
All agents share the same initialization of weights, with the weights being chosen based on a combination of trying to achieve the minimal fit in terms of least-square error to the expected value function, crucially on held-out validation data (to avoid overfitting), and to ensure that the initial fit exhibits desirable properties such as smoothness and monotonicity. The latter I have not formally enforced and instead chose based on visual inspection of the fit, a weakness of the approach.
There may be slight selection bias incurred as a result.

As a result of local fluctuations continuing to be present in policy, as mentioned previously, I also assume that subsequently the agent takes the neural network fit and fits a moderate-order polynomial \emph{on the expected value function} after learning to achieve global smoothness. This helps to enforce the monotonicity of the consumption policy. At present, the model assumes the agent only uses these for considering counterfactuals, such as policy functions and not for learning. That is, explicitly, when computing marginal propensities to consume out of transfers and consumption and savings responses from cross-sectional data, smoothed values are not utilized. Instead, at present, these are solely used for displaying policies within an individual agent and the cross-sectional policy and MPC distributions.\footnote{As an alternative, I considered using a local smoothness criterion or local quadratic variation in the derivative of the fitted value function to minimize perturbations. Unfortunately, this only minimized local perturbations in the fit on the expected value function and not on the policies themselves, which is what I care about when enforcing monotonicity. These instead had to do with local, small fluctuations in the slope of the expected value function in assets relative to the slope of the utility function that violated strict concavity of the value function. I also considered explicitly incorporating the polynomial fit into the learning process, but this led to limited variation in policies over time at low order polynomials and memory issues at higher order polynomials that prevented the fit computationally.}

Agents are initialized at the same set of weight parameters so learning is the sole source of diversity of choices beyond assets.
During the actual simulation, for the neural network, no minibatches or parallel evaluations
are used. I continue to use ADAM for the learning step each period.
\subsection{Parameters Summary}
Key parameters are summarized in \cref{tab:all-params}.
\begin{table}[htbp]
  \centering
  \caption{Key Configuration Parameters}\label{tab:all-params}
  \centering
  \begin{tabularx}{\textwidth}{@{}lXX@{}}
    \toprule
    \textbf{Parameter}              & \textbf{Value}                                  & \textbf{Notes}                                 \\
    \midrule
    \multicolumn{3}{l}{\textit{Simulation Parameters}}
    \\
    \midrule
    number of agents                & 50
    &                                                                                                  \\
    number of time periods          & 50
    \\
    learning rate (simulation)      & \(1.1\times 10^{-3} \)
    &                                                                                                  \\
    optimizer (simulation)          & ADAM
    & less sensitive to learning-rate choice                                                           \\
    polynomial degree               & 5
    & Calibrated, time 0 fit                                                                           \\
    \# evaluation grid for fit      & 50 equally spaced
    &                                                                                                  \\
    max asset level (pre-shock)     & 4.5
    &                                                                                                  \\
    min asset level                 & 0.0
    &                                                                                                  \\
    learning rate decay             & \(\mathcal{O}(t^{-1/2})\)
    & \textcite{Robbins1951}                                                                           \\
    observation period              & quarterly
    &                                                                                                  \\
    learning period                 & quarterly
    &                                                                                                  \\
    \(\beta \)                      & 0.9703 (Q), 0.99
    (M)                             & \textcite{ganongSpendingJobfindingImpacts2024}                                                  \\
    interest rate                   & 0.985\% (Q), 4\%
    (Y)                             & \textcite{ganongSpendingJobfindingImpacts2024}                                                  \\
    initial asset distribution      & piecewise linear inverse CDF,
    Pareto tail                     & SCF 2016, financial wealth
    \\
    UE, UU                          & (0.392, 0.608) (Q) (0.28,
    0.72) (M)                       & \textcite{ganongSpendingJobfindingImpacts2024}                                                  \\
    EE, EU                          & (0.939, 0.0607) (Q)
    (0.972, 0.028) (M)              & \textcite{ganongSpendingJobfindingImpacts2024}                                                  \\
    replacement rate                & 0.472 (\$4,268, 2020 dollars)          & \textcite{ganongSpendingJobfindingImpacts2024} \\
    income value                    & 1.0 (\$9,042, 2020 dollars)
    & \textcite{ganongSpendingJobfindingImpacts2024}                                                   \\
    shock size (fraction of income) & 0.784 (\$7,200, 2020 dollars)
    & \textcite{ganongSpendingJobfindingImpacts2024}                                                   \\
    \midrule
    \multicolumn{3}{l}{\textit{Neural Network Configuration}}
    \\
    \midrule
    size, hidden dimension          & 80
    &                                                                                                  \\
    \# of hidden layers             & 2
    &                                                                                                  \\
    activation                      & Rectified Linear (ReLU)
    &                                                                                                  \\
    \midrule
    \multicolumn{3}{l}{\textit{Grid Configuration}}
    \\
    \midrule
    asset grid size                 & 500 states
    &                                                                                                  \\
    \midrule
    \bottomrule
  \end{tabularx}
\end{table}

\section{Mechanism}\label{sec:mechanism}

\subsection{Experiment with Extreme Shocks}\label{sec:extreme-shocks}
\begin{figure}
  \centering
  \includegraphics[width=1\linewidth]{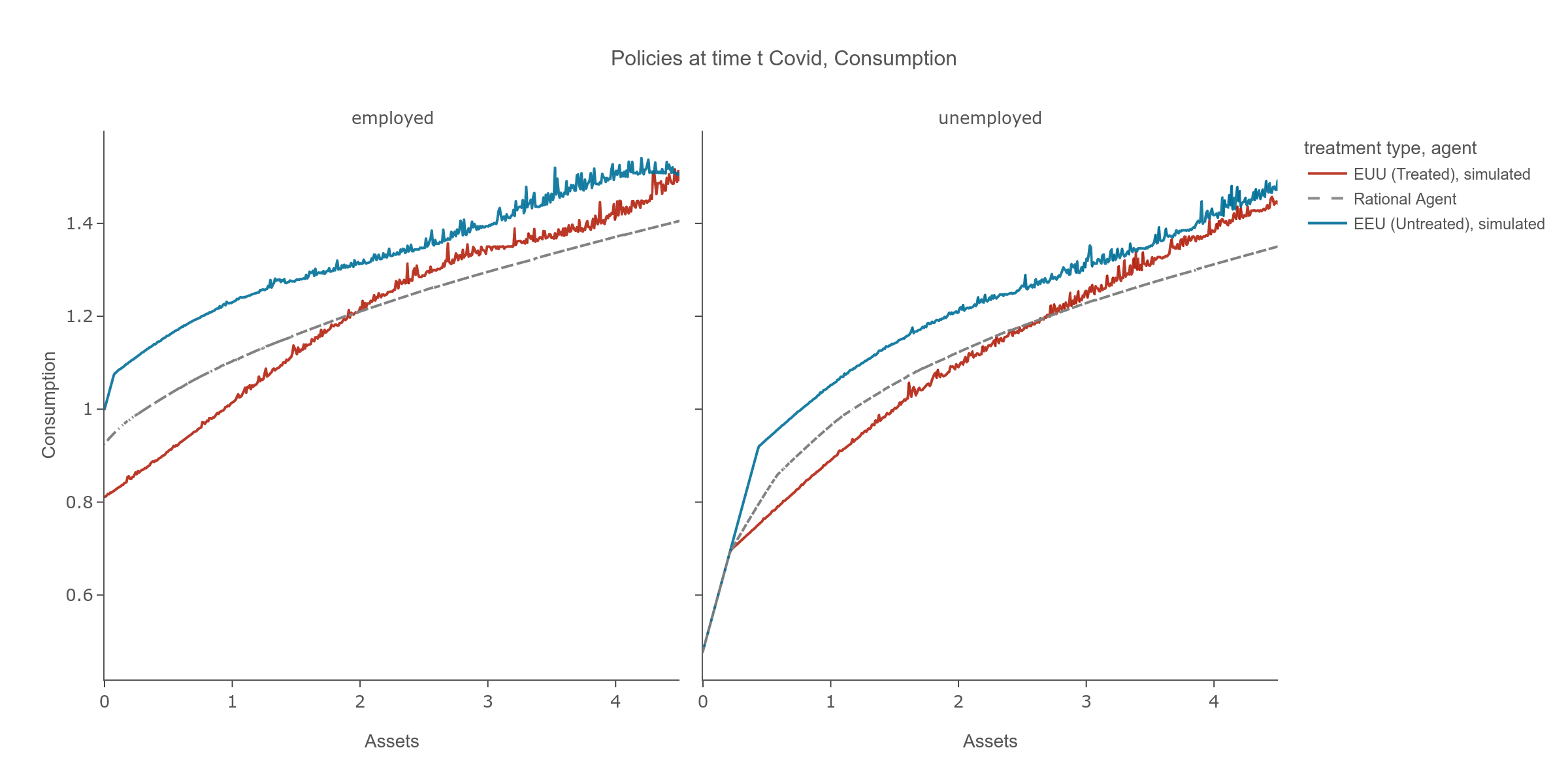}
  \caption[Consumption in Repeated Employment and Unemployment Scenario, 5 Periods]{Consumption policies at \(t = 5\) for the experiment with repeated employment with one agent receiving a series of unemployment realizations and the other receiving a series of employment realizations. The blue line is the agent with unemployment experiences, while the red line is the agent with employment experiences. The dotted gray line is the rational agent's policy. The figure on the left represents the agent's consumption policy as a function of their assets (x-axis) and income while employed \(y_e\), while on the right it represents the agent's consumption policy as a function of their assets and income while unemployed \(y_u\). Value functions use a polynomial fit.}
  \label{fig:experiment-last-period-cons}
\end{figure}

To highlight the mechanism, I conduct stylized experiments with two agents simulated in parallel. Two agents were simulated in parallel, with both agents starting with employment, the same asset level (the median). Then, in the subsequent period, one agent receives a series of unemployment realizations for four periods, while the other receives a series of employment realizations. The intent is to generate opposite extreme experiences in order to better observe
how the impact of employment or unemployment realizations is affecting the agents' policies. Both are plotted relative to the rational benchmark. Figure~\ref{fig:experiment-last-period-cons} shows that for sufficiently high asset levels, both agents consume more than the rational benchmark. However, for low asset levels, the agent with more unemployment experiences exhibits significant scarring, represented
by a shift down in the consumption policy curve. By contrast, the agent with more employment experiences consumes more than the rational benchmark everywhere, an ``anti-scarring'' effect due to too low risk.
By comparison, see \cref{fig:initial-cons-policy} for the initial fit.

\begin{figure}
  \centering
  \includegraphics[width=1\linewidth]{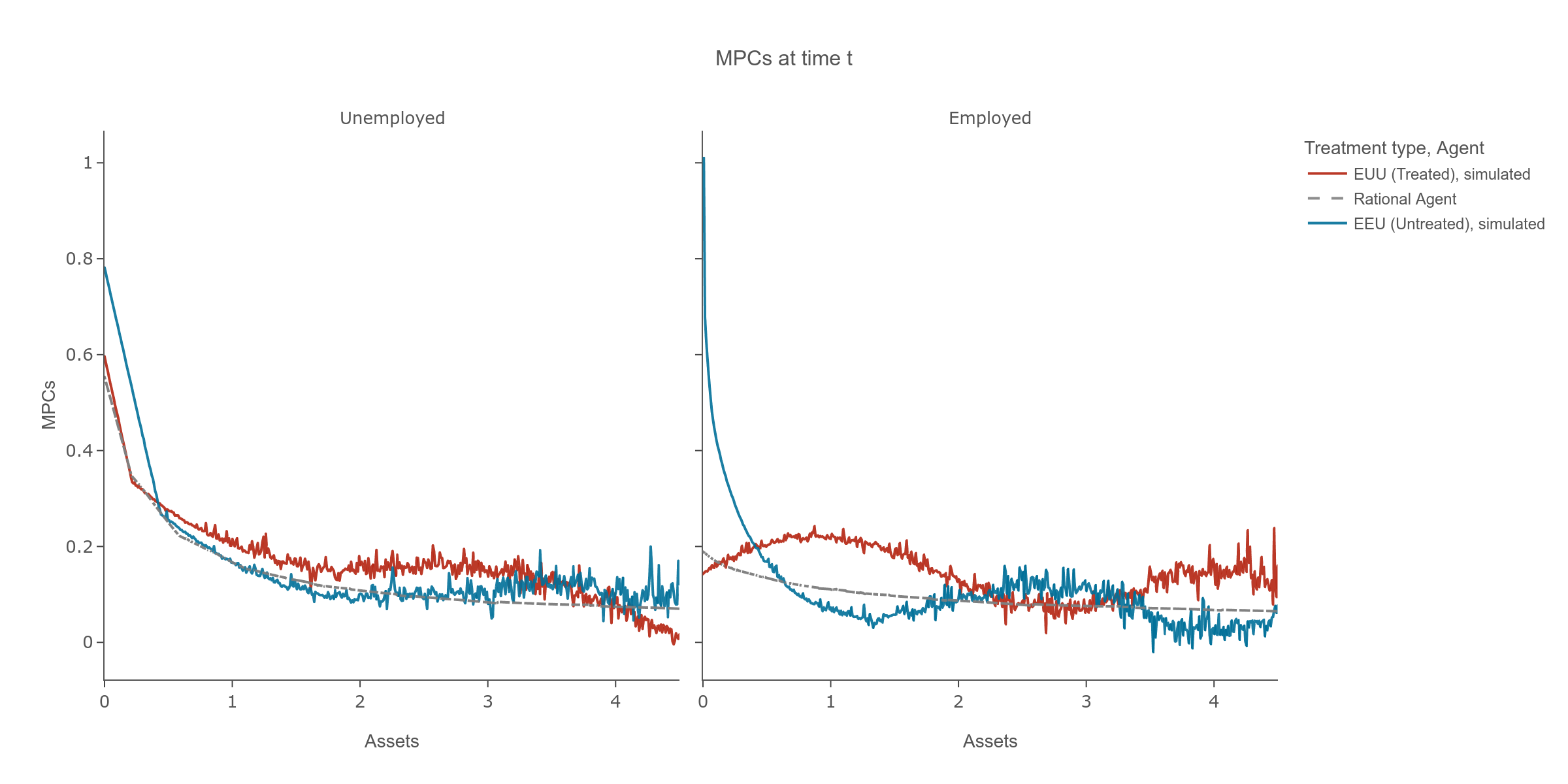}
  \caption[MPCs After 5 Periods Repeated Employment and Unemployment]{MPCs for the experiment with repeated employment and unemployment realizations after \(t =5\) periods, where one agent receives a series of unemployment realizations and the other receives a series of employment realizations. The blue line is the agent with unemployment experiences, while the red line is the agent with employment experiences. The dotted gray line is the rational benchmark's policy. The figure on the left represents the agent's marginal propensity to consume as a function of their assets (x-axis) and income while employed \(y_e\), while on the left it represents the agent's consumption policy as a function of their assets and income while unemployed \(y_u\). This uses the polynomial fit to the value function generating local oscillations from the polynomial.}
  \label{fig:experiment-last-period-mpcs}
\end{figure}

Close to the no-borrowing constraint, an agent with more unemployment experiences
appears to slightly decrease their consumption, while employed, more than the agent with employed experiences does while unemployed. This is a nice sanity check that the value-based mechanism works, since risk aversion and the borrowing constraint imply asymmetrically
larger losses in the value function for the agent with more unemployment experiences under equivalent policies, partially reflected in the curvature of the consumption policy function. By
contrast, the slope of the consumption policy for the agent with more unemployment experiences is generally steeper than the agent with more employment experiences, which is suggestive of a higher marginal
propensity to consume (MPC) for the agent with more unemployment experiences; the agent's MPC is estimated as a local slope between any two points on the consumption policy curve separated by the transfer
quantity. I verify this with a graph of the MPCs as a function of assets in \cref{fig:experiment-last-period-mpcs} and see that in general, for values close to \(0\) assets, the agent receiving more employment experiences has higher MPCs, but for values further away, the agent receiving more unemployment realizations has higher MPCs.

\subsection{How SGD Updates Shape the Expected Value Function}\label{sec:mechanism-theory}

The key mechanism by which the reinforcement learning agent generates both lower consumption and higher MPCs after unemployment experiences operates through the interaction of the temporal difference (TD) error and the neural network's gradient update rule. This subsection develops the intuition for how this occurs.

Consider an agent who experiences an unemployment shock. The TD error,
\begin{align}
  \epsilon(a,y,a', y';\phi) = \beta\Big[\max_{a''}Q(a',y',a'';\phi)
  - \widehat{EV}(y,a';\phi)\Big],
\end{align}
captures the difference between the realized value of savings tomorrow and the agent's prior expectation. An unemployment realization (\(y' = y_u\)) generates a lower realized \(\max_{a''}Q(\cdot)\) than expected, producing a negative TD error. This negative TD error pulls the estimated \(\widehat{EV}\) downward locally via the gradient update rule \cref{eq:learn}.

Crucially, the neural network does not update \(\widehat{EV}\) at a single point. Instead, it adjusts the global parameter vector \(\phi\) in the direction \(\nabla_\phi \widehat{EV}(y,a';\phi)\). For ReLU networks, this gradient-based update has a natural distance-decay property: the magnitude of the update to \(\widehat{EV}\) at any point \(\tilde{a}'\) is larger when \(\tilde{a}'\) is closer to the agent's current savings choice \(a'^*\) and smaller when \(\tilde{a}'\) is far away.\footnote{This distance-decay property is related to the neural tangent kernel literature \parencite{Jacot2018}, which shows that for wide networks, updates at one input point affect nearby inputs more strongly than distant ones, with the rate of decay governed by the architecture.} This extrapolation pattern produces asymmetric effects on the shape of \(\widehat{EV}\):

\begin{enumerate}
  \item \textbf{Slope increases locally} (steepens): because the update pulls \(\widehat{EV}\) down more at the agent's current savings level than at higher savings levels, the function becomes steeper in the neighborhood of the agent's assets.

  \item \textbf{Curvature increases} (becomes more concave): the differential magnitude of updates at nearby versus distant asset levels increases the second derivative of \(\widehat{EV}\).
\end{enumerate}

\begin{figure}[htbp!]
  \centering
  \includegraphics[width=1\linewidth]{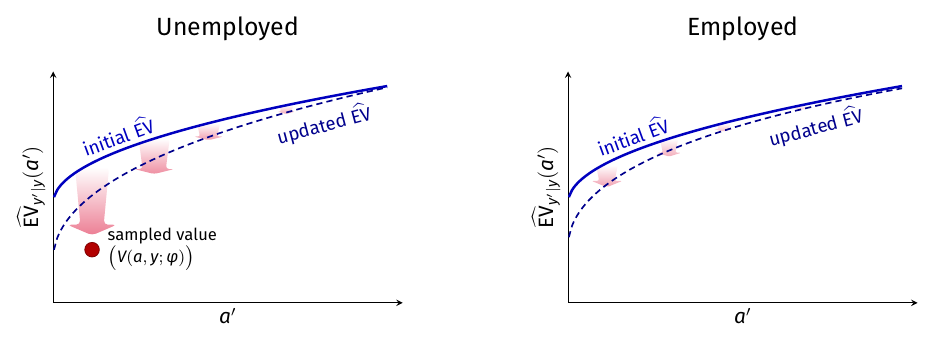}
  \caption[How SGD updates shape the expected value function]{Neural network gradient updates produce larger changes to \(\widehat{EV}\) at savings levels closer to the agent's current position. This increases the slope and curvature of \(\widehat{EV}\), which in turn decreases average consumption and increases MPCs out of a one-time transfer.}
  \label{fig:conceptual-ev-update}
\end{figure}

\Cref{fig:conceptual-ev-update} illustrates this mechanism. The initial \(\widehat{EV}\) (blue solid) shifts to the updated \(\widehat{EV}\) (blue dashed) after a single unemployment realization. Gradient arrows show the direction and magnitude of updates, with larger arrows at asset levels closer to the agent's current position.

From the agent's Euler equation, a steeper \(\widehat{EV}\) (higher \(\widehat{EV}'(a')\)) implies the agent values tomorrow's savings more, leading to lower current consumption---a scarring effect. Simultaneously, a more concave \(\widehat{EV}\) (more negative \(\widehat{EV}''(a')\)) implies the marginal return to saving declines more quickly with asset level, which raises MPCs. For high-asset agents, consumption may actually increase because the gradient update decays sufficiently that the net effect on \(\widehat{EV}'\) at their asset level is negative, consistent with the patterns observed in \cref{fig:experiment-last-period-cons}.

\subsection{Simulated Counterpart: EV Before and After Updates}\label{sec:ev-counterpart}

\begin{figure}[htbp!]
  \centering
  \includegraphics[width=0.85\linewidth]{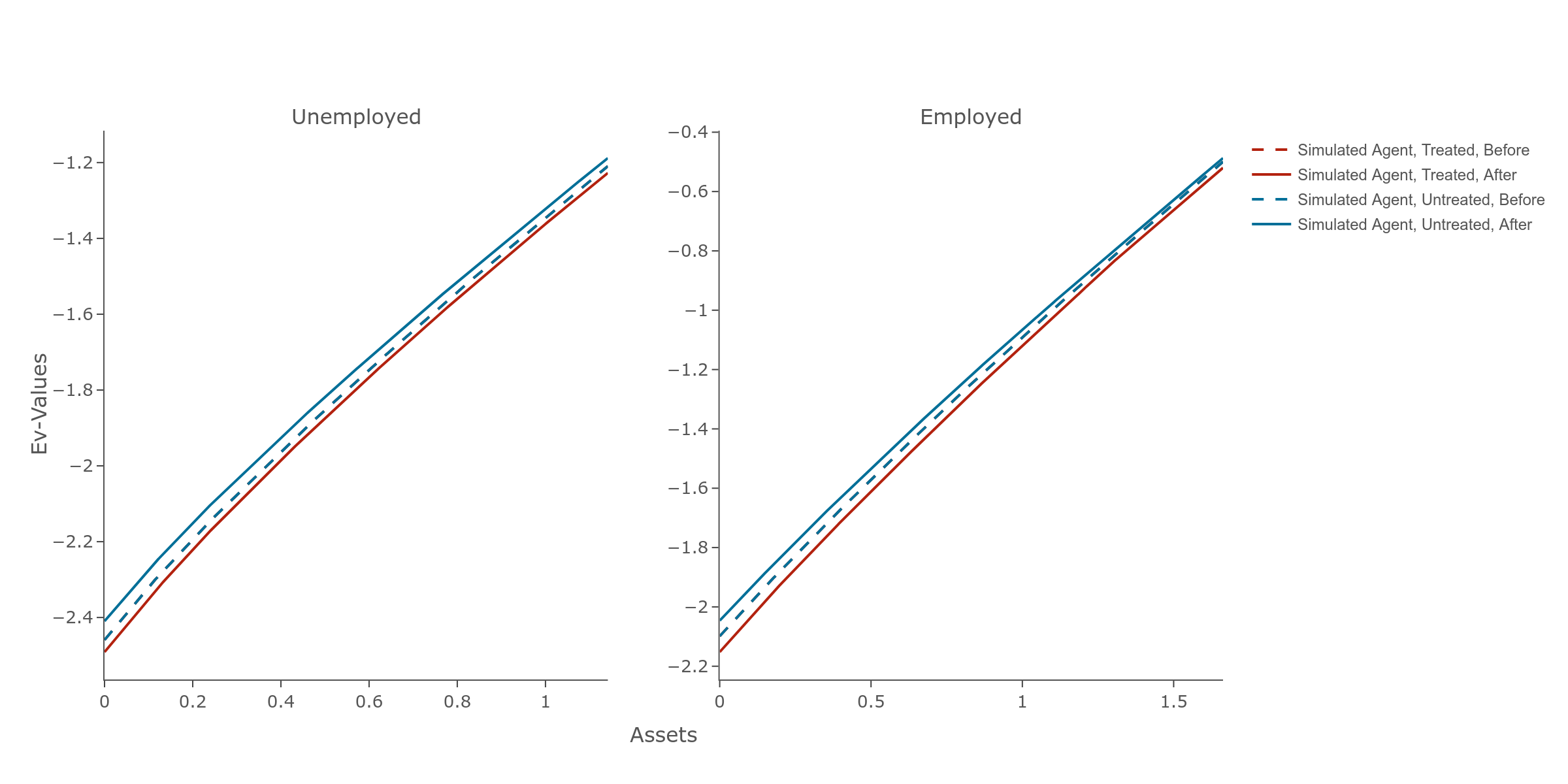}
  \caption[\(\widehat{EV}\) before and after a single update]{\(\widehat{EV}\) evaluated at the agent's current asset levels before (blue) and after (red) a single gradient update at \(t=0\) following 5 periods of unemployment. The x-axis is next-period assets \(a'\). The red curve steepens (lower consumption, higher savings) and becomes more concave (higher MPCs).}
  \label{fig:ev-before-after-t0}
\end{figure}

\Cref{fig:ev-before-after-t0} confirms the theoretical prediction using simulated \(\widehat{EV}\) functions. The figure shows \(\widehat{EV}\) evaluated at the agent's current asset levels before (blue) and after (red) a single parameter update following an unemployment realization. The red curve steepens relative to blue, indicating that saving more for tomorrow has become more valuable (higher slope), and the red curve becomes more concave, indicating that the marginal return to saving declines more quickly with asset level (higher curvature). Both patterns match the theoretical predictions from \cref{sec:mechanism-theory}.

\begin{figure}[htbp!]
  \centering
  \includegraphics[width=0.85\linewidth]{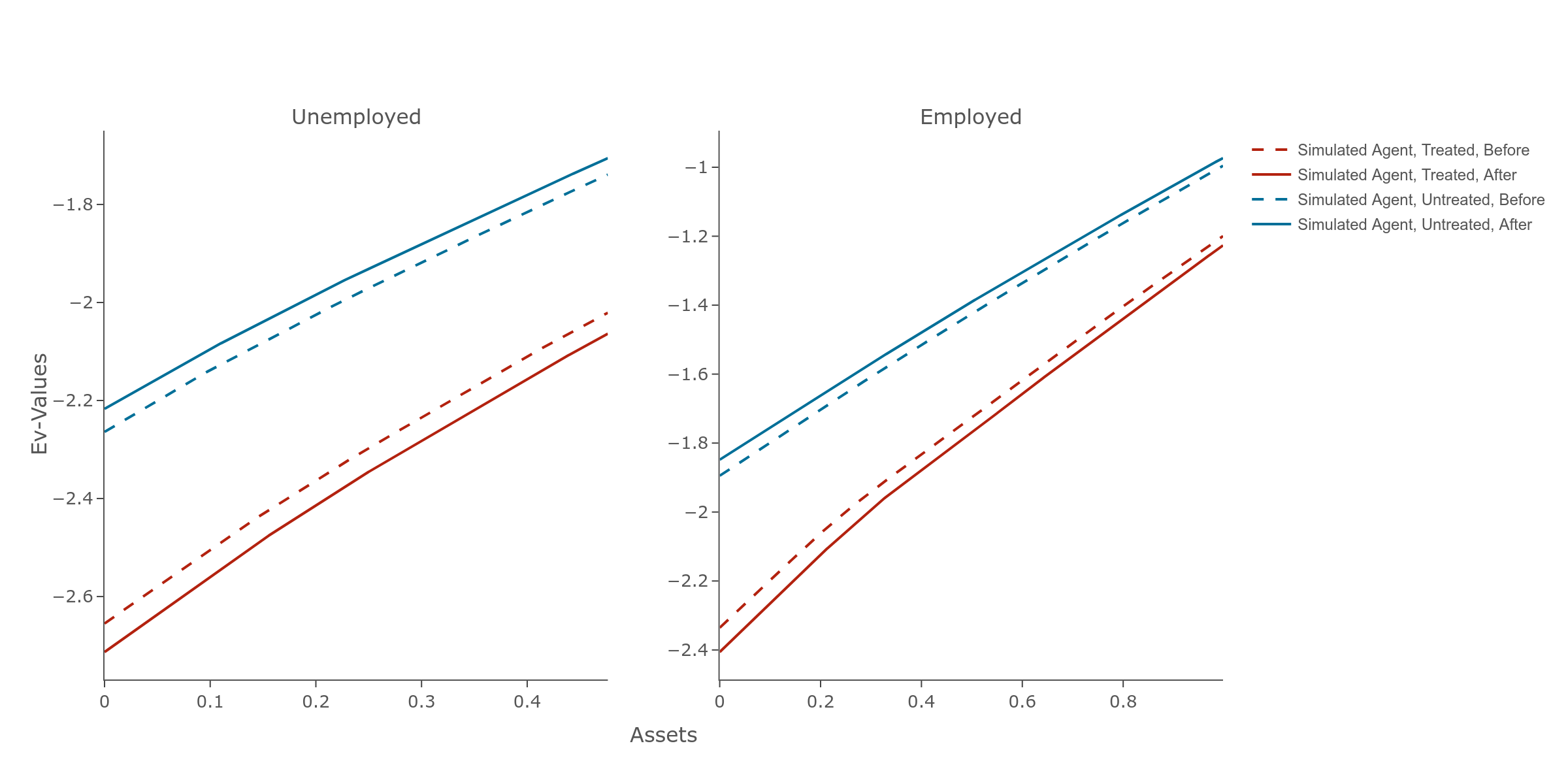}
  \caption[\(\widehat{EV}\) before and after the 4th update]{\(\widehat{EV}\) before (blue) and after (red) the 4th gradient update. The pattern of steepening and increased curvature persists, consistent with the cumulative mechanism.}
  \label{fig:ev-before-after-t4}
\end{figure}

\Cref{fig:ev-before-after-t4} shows the same comparison on the 4th update, demonstrating that the pattern persists across multiple learning steps.

\subsection{Illustrative Example: Polynomial EV}\label{sec:polynomial-mechanism}

To build further intuition, consider a simplified setting with a single asset \(a\), fixed income \(\overline{y}=1\), \(R=1.03\), \(\beta=0.96\), log utility, and no borrowing constraint. Suppose the agent's estimated expected value function takes the simple polynomial form
\begin{equation}
  \widehat{EV}(a') = g_0 + g_1 a' + g_2 {a'}^2,
  \label{eq:polynomial-ev}
\end{equation}
where monotonicity requires \(g_1 > 0\) and strict concavity requires \(g_2 < 0\).

From the agent's first-order condition, \(u'(c) = \beta \, \widehat{EV}'(a')\), which for log utility gives
\[
  \frac{1}{c} = \beta (g_1 + 2 g_2 a').
\]
Applying the endogenous grid method, one can recover the consumption function \(c(a)\) and numerically compute MPCs.

\begin{figure}[htbp!]
  \centering
  \includegraphics[width=0.65\linewidth]{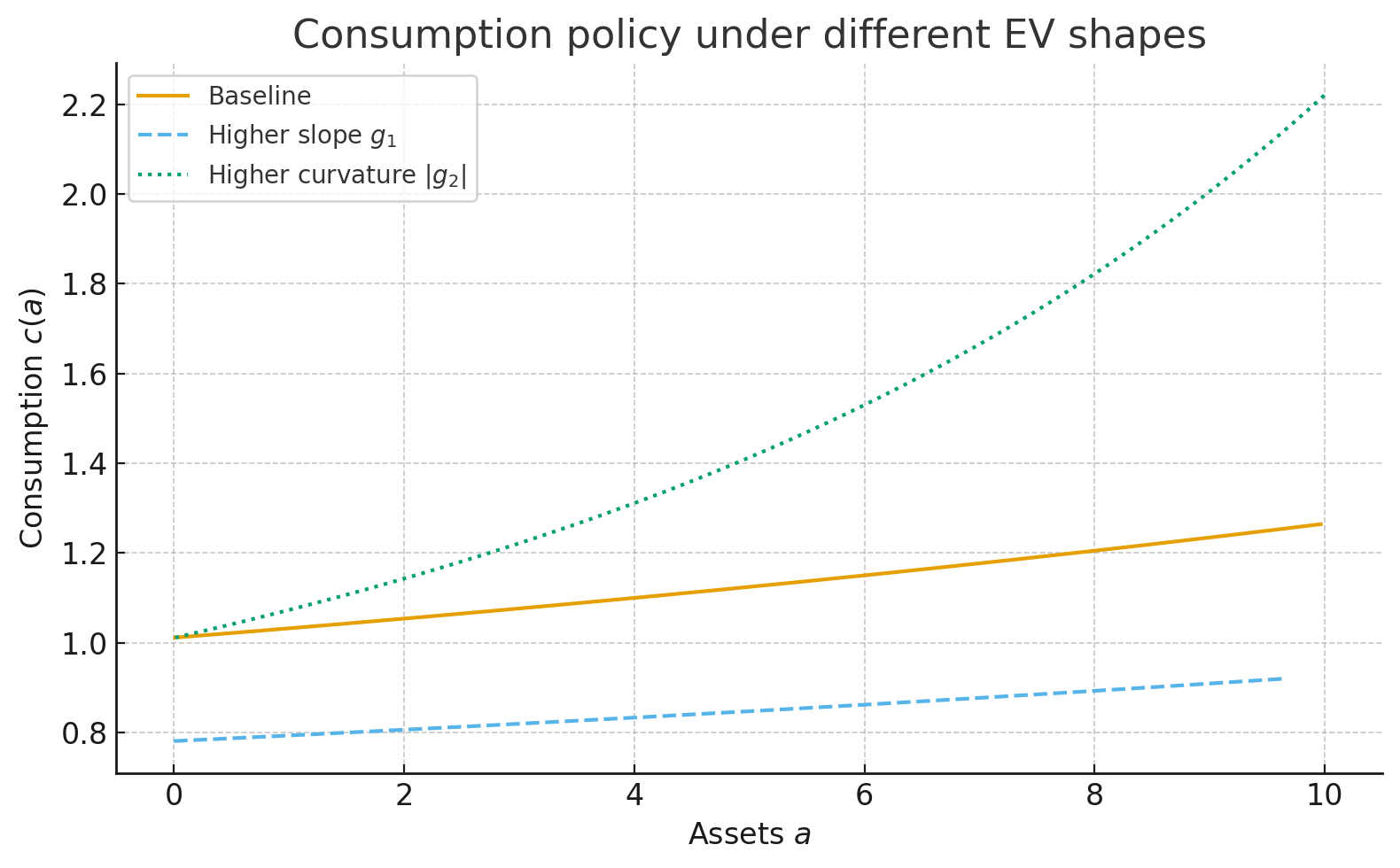}
  \caption[Consumption policy under different EV shapes]{Consumption policy as a function of assets for different values of slope (\(g_1\)) and curvature (\(g_2\)) parameters in the polynomial \(\widehat{EV}\) specification \cref{eq:polynomial-ev}. Higher slope reduces consumption everywhere (scarring). Higher curvature raises consumption at high asset values.}
  \label{fig:polynomial-consumption}
\end{figure}

\begin{figure}[htbp!]
  \centering
  \includegraphics[width=0.65\linewidth]{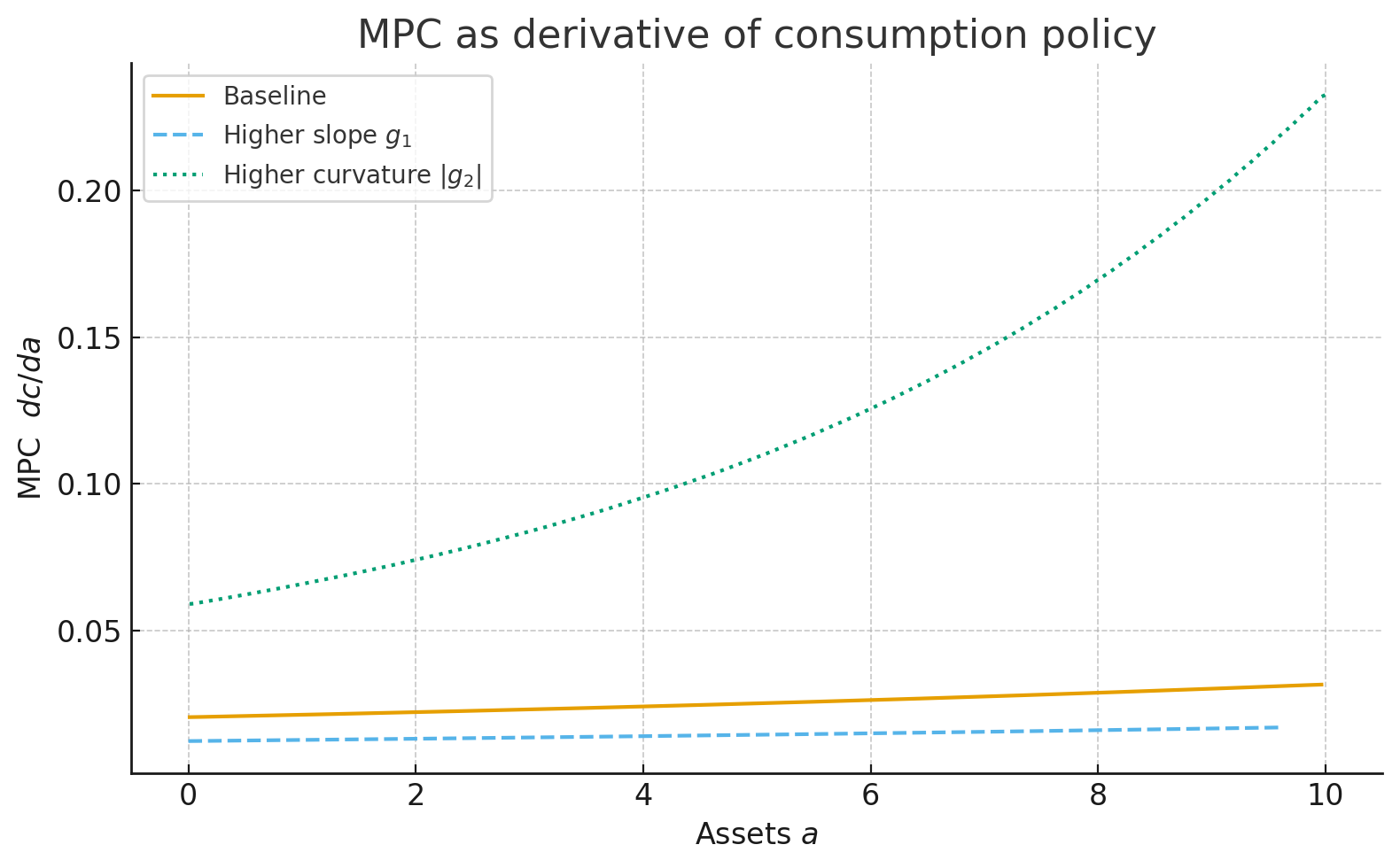}
  \caption[MPC policy under different EV shapes]{MPCs as a function of assets for different values of slope (\(g_1\)) and curvature (\(g_2\)) in the polynomial \(\widehat{EV}\). Higher curvature substantially increases MPCs everywhere. Higher slope slightly decreases MPCs.}
  \label{fig:polynomial-mpc}
\end{figure}

\Cref{fig:polynomial-consumption,fig:polynomial-mpc} show the comparative statics of changing \(g_1\) and \(g_2\) separately:

\begin{itemize}
  \item \textbf{Increasing the slope} (\(g_1\) from 1 to 1.3, steeper EV): saving for tomorrow becomes more valuable everywhere, so consumption decreases globally---a scarring effect.

  \item \textbf{Increasing the curvature} (\(g_2\) from \(-0.02\) to \(-0.05\), more concave EV): the marginal return to saving at high asset levels falls disproportionately, raising MPCs substantially while having a relatively small effect on average consumption levels.

  \item \textbf{Both together}: consumption falls and MPCs rise, precisely matching the joint pattern observed in the full neural network simulations and in the empirical data.
\end{itemize}

This simplified example makes transparent the separate roles of slope and curvature in generating the joint consumption-MPC pattern. The full neural network model combines both effects through the gradient update mechanism described in \cref{sec:mechanism-theory}, where unemployment shocks simultaneously increase both the slope and curvature of \(\widehat{EV}\).

\section{Experiments}
From our initial fit, I then let the agent learn for 50 periods, or about 12.5 years. I then repeat the analysis of
\textcite{ganongSpendingJobfindingImpacts2024} and
\textcite{Malmendier2024}.
\subsection{Marginal Propensities to Consume}
I repeat the primary setup of \textcite{ganongSpendingJobfindingImpacts2024} with minor modifications.

\paragraph{Empirical Specification}
The primary identification strategy of \textcite{ganongSpendingJobfindingImpacts2024} employs a difference-in-differences design exploiting administrative delays in unemployment benefit processing. The treated group consists of workers who became unemployed at the end of March 2020 and received benefits in April, while the control group includes workers who also became unemployed in March but experienced processing delays and did not receive benefits until June. The two-stage specification is:

First stage:
\[
  y_{i,t} = \alpha + \beta \text{Post}_t \times \text{Treat}_i + \text{Treat}_i + \text{Post}_t + \varepsilon_{i,t}
\]

Second stage:
\[
  c_{i,t} = \phi + \text{MPC} \times \hat{y}_{i,t} + \text{Treat}_i + \text{Post}_t + \varepsilon_{i,t},
\]
where \(t \in \{\text{March}, \text{May}\}\), \(\text{Treat}_i\) indicates receipt of benefits in April, and \(\text{Post}_t\) indicates May observations. The coefficient MPC captures the marginal propensity to consume out of unemployment benefits.

To examine heterogeneity by liquidity status, households are split by their 2018 median liquidity buffer, defined as \((\text{checking balance}_{it} - 0.5 \times \text{spending}_{it})/\text{spending}_{it}\). This pre-pandemic measure avoids endogeneity concerns while capturing persistent differences in savings behavior. The paper finds MPCs of 0.53 for low-liquidity households versus 0.29 for high-liquidity households, with effects persisting even after benefit expiration.

Crucially, because I do not have unobserved characteristics and have direct access to asking agents about what they would have done if they had received a transfer at any given time, I do not need to repeat most of the regression of \textcite{ganongSpendingJobfindingImpacts2024}, which aims to control for unobserved heterogeneity and indirectly estimate group average MPCs. Instead, I  ``freeze'' agents (so that they do not learn temporarily from new experiences) and directly provide agents with and without a transfer at time \(t\) and see how their consumption choices would vary given their value functions and policies.

Crucially, I still split agents at time 0 by whether they are above or below the
median asset holding at time 0 in the simulation. Agents who fall
above are called \emph{high liquidity} while those below are \emph{low liquidity}. Because there is no exact analog of bank balance sheets, assets are used as the measure instead.
I then compute group average MPCs out of transfers eight periods
later, the same duration as the beginning of
\textcite{ganongSpendingJobfindingImpacts2024} until the payments.
MPCs are computed by shifting agents' assets by the size of the COVID
stimulus UI expansion, showing agents both asset levels at the same
income level and evaluating how consumption changed relative to how
assets changed.
Because \textcite{ganongSpendingJobfindingImpacts2024} is focused on
MPCs for the unemployed, I only examine agents while they are unemployed. I consider all MPCs for any agent between quarters eight and nine, which covers the same duration of four months of FPUC transfers from \textcite{ganongSpendingJobfindingImpacts2024}.
\subsection{Results for Marginal Propensities to Consume}
In \cref{tab:ganong2024}, I show the results of the MPC
calculation for agents at eight periods after the initial period, when unemployed. In \cref{tab:ganongmultiperiod}, I repeat this exercise but now with agents split up at different points in the simulation by liquidity level.
\begin{table}[!htbp]
  \centering
  \begin{threeparttable}
    \caption{Model versus Empirical Marginal Propensities to Consume (MPCs)}
    \begin{tabular}{lccc}
      \toprule
      Liquidity Type & Model MPC & Empirical MPC\tnote{a} & Rational MPC Estimates\tnote{b} \\
      \midrule
      Low            & 0.501     & 0.53                   & 0.03 (1-asset), 0.16 (2-asset) \\
      High           & 0.343     & 0.29                   & 0.03 (1-asset), 0.16 (2-asset) \\
      \bottomrule
    \end{tabular}
    \label{tab:ganong2024}
    \begin{tablenotes}[flushleft]
      \small
    \item \textit{Notes:} This table compares MPCs for unemployed agents, categorized by their initial asset level (liquidity) relative to the median. The MPC is measured for agents unemployed between periods \(t=8\) and \(t=9\).
    \item[a] Empirical values and methodology replicate the ``preferred specification'' from \textcite{ganongSpendingJobfindingImpacts2024}.
    \item[b] Rational MPC estimates from \textcite{Kaplan2022}: 0.03 for the one-asset model and 0.16 for the two-asset (liquid + illiquid) model, representing quarterly MPCs out of an unexpected windfall for borrowing-unconstrained households.
    \end{tablenotes}
  \end{threeparttable}

\end{table}
\begin{table}[!htbp]
  \begin{adjustbox}{width=\textwidth,center}
    \centering
    \begin{threeparttable}
      \caption{Model MPCs by Periods Experienced and Liquidity Type}
      \label{tab:mpc_dynamics}
      \small
      \begin{tabular}{@{}lcccccccccc@{}}
        \toprule
        & Empirical\tnote{a} & \multicolumn{9}{c}{Model: Periods Experienced in Simulation Prior to Classification (quarters)} \\
        \cmidrule(lr){2-2} \cmidrule(lr){3-11}
        Liquidity Type & Benchmark & \(0\) & \(5\) & \(10\) & \(15\) & \(20\) & \(25\) & \(30\) & \(35\) & \(40\) \\
        \midrule
        Low  & 0.530 & \textbf{0.501}\(^{*}\) & \textbf{0.589}\(^{***}\) & \textbf{0.585}\(^{***}\) & 0.528 & 0.571 & 0.531 & 0.516 & 0.454 & 0.559 \\
        High & 0.290 & 0.343 & 0.334 & 0.400 & 0.470 & 0.520 & 0.492 & 0.522 & 0.468 & 0.587 \\
        \midrule
        Difference & 0.240 & \textbf{0.158}\(^{*}\) & \textbf{0.255}\(^{***}\) & \textbf{0.185}\(^{***}\) & 0.058 & 0.051 & 0.038 & -0.007 & -0.014 & -0.028 \\
        & & (2.15) & (4.44) & (5.35) & (0.77) & (0.65) & (0.50) & (-0.13) & (-0.19) & (-0.97) \\
        \bottomrule
      \end{tabular}
      \label{tab:ganongmultiperiod}
      \begin{tablenotes}[flushleft]
        \small
      \item \textit{Notes:} *** \(p<0.01\), ** \(p<0.05\), * \(p<0.10\). Null hypothesis is no difference between low and high.\\
        MPCs measured 8 quarters after liquidity classification, including for empirical. Differences show low minus high liquidity MPCs with t-statistics in parentheses (Welch's t-test with unequal variances).
      \item[a] Empirical values and methodology replicate the ``preferred specification'' from \textcite{ganongSpendingJobfindingImpacts2024}.
      \end{tablenotes}
    \end{threeparttable}
  \end{adjustbox}
\end{table}

\subsection{Scarring}
I repeat the analysis of \textcite{Malmendier2024}
by computing a statistic for each agent at each time period that
varies between 0 and 1 based on how recent past unemployment
experiences were.
\paragraph{Empirical Specification}
An index of 0 indicates no unemployment experiences in an
agent's past, while if an agent only experienced unemployment for all
past periods, they would have an index of 1. Experiences further back in the past are down-weighted in how many time periods ago they occurred.
The unemployment index is based on indicators for unemployment weighted by
\[
  w(t,k)=\frac{(t-k)}{\sum_{k=2}^{t-1}(t-k)}.
\]
The numerator represents the relative importance while the
denominator normalizes the weights to sum to 1.
\(t\) is the current time period and \(k\) is the number of time
periods ago the unemployment experience
occurred. \textcite{Malmendier2024} begin with
time period 1. The paper excludes the most recent two time periods
(today and yesterday) to exclude possible shorter term,
non-experience-based effects, which I also follow, though it is important to note that while I operate on quarterly frequency data, \textcite{Malmendier2024} work with biennial frequency data. Time periods in the authors' paper are calibrated to
represent quarterly data in simulation and biennial samples of annual data in the data
from PSID, with a maximum of 18 years of data from PSID (1999--2017; 9 periods), and 50 years in a lifecycle simulation (200 quarters).
The current value of the index for an agent is given by \(p_{it}\) defined as
\[
  p_{it} = \sum_{k=2}^{t-1}w(t,k) W_{t-k},
\]
where \(W_{t-k}\) is an indicator for whether an agent was unemployed
at time \(t-k\).
The specification for the regression I utilize, with quarterly
frequency, is based on the regression:
\[
C_{it}=\alpha + \phi UE_{{pers},{i,t}} + \gamma' x_{it} +\varepsilon_{it}\]
where \(C_{it}\) is the consumption of agent \(i\) at time \(t\),
\(\phi \) is the coefficient on the unemployment experience index,
\(UE_{pers,i,t}\), and \(x_{it}\) are controls that consist of
\((a_{it}, y_{it})\) asset income pairs for us.

Note that the original paper uses the full specification
\[
  C_{i t}=\alpha+\psi U E P_{i t}+\beta U E_{i t}+\gamma^{\prime}
  \mathbf{x}_{i t}+\eta_t+\delta U_{s t}+\varsigma_s+v_i+\epsilon_{i t},
\]
where \(x_{it}\) is a series of controls including demographic,
income, and include: \(\log(y_t)\), \(\log(\log(y_t))\), \(\log(y_{t-1})\),
\(\log(\log(y_{t-1}))\), \(\log(a_t)\), \(\log(\log(a_t))\) controls
(shifted first from their minimal value, so that log is
well-defined); \(UEP_{it}\) the personal measure of unemployment;
\(\eta_t\) a time fixed effect; \(v_i\) an agent fixed effect;
\(\varsigma{s}\) a state fixed effect; \(U_{st}\) the state
unemployment level at the present time; and \(UE_{it}\) the
macroeconomic unemployment index, constructed identically to how I
constructed the individual level index, except using the state-level
unemployment percentage for \(W_t\) at time \(t\). Given that I do not
have confounding present in the simulation beyond asset and income
experiences, as agents all begin with the same weighting, I do not
include these additional controls in the regression. Similarly,
because I do not have extreme skewness in wealth, I control for
wealth only via the asset level \(a_{it}\) and do not include
\(\log(a_{it}), \log(\log(a_{it}))\) controls or income controls.
Finally, \textcite{Malmendier2024} truncate agents at the tenth percentile of wealth and the ninetieth percentile of total income. I similarly do not do this.
\subsection{Results for Scarring}
\newcommand{\simregval}{-0.0378}
In \cref{tab:malmendier2024}, I show the results of the
regression of consumption on the unemployment experience index, as
well as the average MPCs for agents with different unemployment
experience indices. With approximately 4,700 observations (over multiple time periods), I find about a \(\simregval\)\% drop in consumption for a 1\% increase in the unemployment index.
This effect is small but significant at the 1\% level; it is, however, about an order of magnitude smaller than \citeauthor{Malmendier2024}'s estimate at the biennial frequency of \( - 0.280\). Given their difference in frequency, they may not be directly comparable in magnitude. When not controlling for assets, I find a larger drop in consumption of around 0.0884\% at the 1\% significance level. This suggests that both assets and unemployment experiences play an important role in determining consumption changes.
\begin{table}[!htbp]\centering
  \caption[Regression of unemployment experiences on consumption]{Regression of personal unemployment experience on consumption with and without controls for asset levels.}
  \begin{tabular}{@{\extracolsep{5pt}}lcc}
    \\[-1.8ex]\toprule
    \\[-1.8ex] & \multicolumn{2}{c}{\textit{Dependent variable:}} \\
    \cline{2-3}
    \\[-1.8ex] & \multicolumn{2}{c}{consumption} \\
    \\[-1.8ex] & (1) & (2)\\
    \midrule
    \\[-1.8ex]
    UEP\(_t\)                  & \(-\)0.0884\(^{***}\)  & \simregval\(^{***}\) \\
    & (\(-\)0.108, \(-\)0.069) & (\(-\)0.052, \(-\)0.024) \\
    &                    &                   \\
    assets\(_t\)               &                    & 0.1265\(^{***}\)    \\
    &                    & (0.123, 0.130)    \\
    &                    &                   \\
    income\(_t\)               & 1.0229\(^{***}\)     & 0.9869\(^{***}\)    \\
    & (1.020, 1.026)     & (0.985, 0.989)    \\
    &                    &                   \\
    \midrule
    \\[-1.8ex]
    Observations             & 4,692              & 4,692             \\
    R\(^{2}\)                  & 0.658              & 0.835             \\
    Adjusted R\(^{2}\)         & 0.658              & 0.835             \\
    Log Likelihood           & 5,173.100          & 6,878.400         \\
    AIC                      & \(-\)10,340.000      & \(-\)13,750.000     \\
    BIC                      & \(-\)10,330.000      & \(-\)13,730.000     \\
    F Statistic              & 9,015\(^{***}\) (df = 1; 4690) & 11,830\(^{***}\) (df = 2; 4689) \\
    \bottomrule
    \\[-1.8ex]
    \textit{Note:}           & \multicolumn{2}{r}{*** \(p<0.01\), ** \(p<0.05\), * \(p<0.10\)} \\
    & \multicolumn{2}{r}{95\% confidence intervals in parentheses} \\
  \end{tabular}
  \label{tab:malmendier2024}
\end{table}

\subsection{Experiments Summary}
When I run the simulation, I get results qualitatively consistent with both sets of facts. In particular, after 50 periods of running the simulation, I appear to get overall higher consumption levels and higher MPCs than the rational case over 50 time periods. But once selecting for past unemployment experiences, I get evidence of scarring effects.

\section{Discussion}
The agents' MPCs quantitatively and qualitatively replicate the findings of \textcite{ganongSpendingJobfindingImpacts2024}. For \textcite{Malmendier2024}, however, I find only broad qualitative agreement with simulation-based estimates of effect size generally about an order of magnitude smaller than the empirical effect size. However, one important thing to note is that the simulation generated data and \textcite{Malmendier2024} occur at different frequencies, with the simulation's data being quarterly and \citeauthor{Malmendier2024}'s being biennial, representing longer durations and, hence, more time accumulated in the learning process.
Unlike the higher perceived probability of unemployment like \textcite{Malmendier2024}, I am able to generate both higher MPCs and lower average consumption levels. See \cref{app:pessimism} for further details and a simulation result that shows that both MPCs and average consumption levels decline under the current parameterization when solely increasing probabilities of unemployment.

\subsection{Comparison with Pessimistic Beliefs}\label{sec:rl-vs-pessimism}

To sharpen the distinction between the reinforcement learning mechanism and standard belief-based learning, I compare the model's predictions against a ``pessimistic'' agent who updates their beliefs about unemployment probabilities, as in \textcite{Malmendier2024}'s anticipated utility framework. Specifically, I take an agent who perceives 1.5 times the true \(UU\) and \(EU\) transition probabilities (renormalized), and solve the Bellman equation exactly given these pessimistic beliefs. Full results are shown in \cref{app:pessimism}.

\Cref{tab:rl-vs-pessimism} summarizes the directional predictions. The RL agent matches both empirical patterns: lower consumption (scarring) and higher MPCs. The pessimistic agent matches the consumption direction (lower) but generates lower MPCs, opposite to the empirical pattern.

\begin{table}[htbp!]
  \centering
  \caption{Directional Comparison: RL vs.\ Pessimistic Beliefs vs.\ Rational Expectations}\label{tab:rl-vs-pessimism}
  \begin{tabular}{lcc}
    \toprule
    \textbf{Model} & \textbf{Consumption} & \textbf{MPC} \\
    \midrule
    Empirical & \(\downarrow\) & \(\uparrow\) \\
    RL Agent & \(\downarrow\) \(\checkmark\) & \(\uparrow\) \(\checkmark\) \\
    Pessimistic (AU) & \(\downarrow\) \(\checkmark\) & \(\downarrow\) \(\times\) \\
    Rational (benchmark) & \(-\) & \(-\) \\
    \bottomrule
  \end{tabular}
\end{table}

The intuition for why pessimistic beliefs reduce MPCs is as follows. Under anticipated utility, the agent who believes unemployment is more likely solves a Bellman equation to a fixed point given their pessimistic beliefs. This generates a globally lower consumption policy everywhere due to higher precautionary savings. Because the agent exhibits prudence, a lower consumption policy everywhere in turn leads to a lower local slope of the consumption policy and lower MPCs per asset level. That is, pessimism shifts the entire consumption function down uniformly, flattening it, which mechanically reduces the difference in consumption between adjacent asset levels---precisely the MPC.

By contrast, the RL agent's mechanism operates through the \emph{shape} of the estimated \(\widehat{EV}\) function. As discussed in \cref{sec:mechanism-theory}, unemployment shocks increase both the slope (generating scarring) and the curvature (generating higher MPCs) of \(\widehat{EV}\), allowing the model to match both empirical directions simultaneously.

\section{Conclusion}
In this work, I show that agents utilizing reinforcement learning replicate the key findings of
\textcite{ganongSpendingJobfindingImpacts2024} and
\textcite{Malmendier2024} in a model-free setting.
This paper's learning-based mechanism, deep reinforcement learning, is similar to
\textcite{Malmendier2024} in that it is experience-based, but differs in that it explicitly utilizes local ``surprises'' in utility to have an agent adjust their behavior, without having an explicit probability distribution that is updated separately from the agents' value function. This differing mechanism enables us to both generate higher MPCs and lower average consumption levels, something not possible when an agent's perceived probability of transitioning to unemployment occurs. Instead, the expected value function is the only estimator that the agent uses. The neural network estimator of the conditional expected value  does not assume anything about the underlying income distribution.
As a result, I do not need prior knowledge of a particular form of the distribution, such as a two-state Markov switching process.
Regardless of the form, the agent will be able to update their expected value function, using this lower-dimensional statistic to guide their behavior.
The practical trade-off is that without assuming a model, I also cannot extract clear beliefs on the part of the agent over future unemployment probabilities, limiting my ability to make precise statements about the causes of agents' behavior and how their beliefs are changing over time, as well as preventing us from using belief surveys to calibrate the agent from the data.

This is a limitation that can be addressed in future work, by either including a general model-based reinforcement learning scheme or successor representation where an agent learns some flexible approximation of their probability of reaching different future states over time, using proxies for an agent's beliefs in the form of attention-based mechanisms that tell us what information the agent
learns to view most saliently, or by using a model-free
distributional or Bayesian reinforcement learning approach, where the agent includes a distribution over probabilities of future utility realizations/value functions they may face.
A particularly promising direction is to recover and discipline the value function model via the implicit beliefs embedded in learned value functions, following recent advances in model-based reinforcement learning \parencite{Hafner2025} and work on extracting implicit beliefs from model-free agents \parencite{Richens2025}.

Another limitation of this work is that, at present, while I test against the two key papers, I do not fully qualitatively match both results, nor test against empirical data.
This, too, presents a future direction of inquiry.
In this paper, I do not touch on theoretical results on convergence or dynamics, which would be interesting to explore.
I focus solely on the behavior of a single agent. A natural question would be what economic equilibria look like when many agents using reinforcement learning interact, such as in an Aiyagari model, what equilibria are learnable, and how they differ from the rational expectations equilibrium, if it is not learned.

Finally, the assumption of polynomial smoothed value functions can possibly be relaxed by instead using a policy function to interpolate choices across policies, such as via an actor-critic approach. This also presents a clearer separation between how agents judge the value of different choices, at a small set of values and how they decide, in a potentially suboptimal manner, which actions to take. At present, both mechanisms are combined in the same neural network, which makes it difficult to disentangle the separate motivations.

\appendix
\renewcommand\thefigure{A\arabic{figure}} 
\renewcommand\thetable{A\arabic{table}}
\renewcommand\theequation{A\arabic{equation}}
\renewcommand\thesubsection{A.\arabic{subsection}}
\setcounter{figure}{0}
\setcounter{table}{0}
\setcounter{equation}{0}
\setcounter{subsection}{0}
\newappendix{Appendix:\@ Supplementary Material}
\subsection{Environment}
Formally, my setting can be defined as the Markov Decision Process
(MDP) given by the tuple \(\mathcal{M} = (\mathcal{S}, \mathcal{A},
P_{yy'}, u) \), where the state space is given by all possible asset,
income pairs, \(\mathcal{S} = {\{a,y\}}_{a \in \mathbb{R}^{++}, y \in
\{y_{u},y_{e}\}} \) and \(\mathcal{A} = \qty{c}_{c \in \mathbb{R}^{++}} \).
Note that because of my special problem structure, the standard
\(P(s,s') \) for  states \(s \in \mathcal{S} \) reduces to \(P_{yy'}
\) as there is no exogenous uncertainty regarding how \(a'\) evolves.
Similarly, \(a'\) is determined exactly by \(c\) and so can be omitted from the action space. The ``reward'' \(R\) depends solely on the agent's choice \(c\) and not on the surrounding state \(s\), and is equivalent to \(u(c)=\log(c) \).
\subsection{Unsmoothed Setting}\label{app:unsmoothed}
\begin{figure}[htbp!]
  \includegraphics[width=1\linewidth]{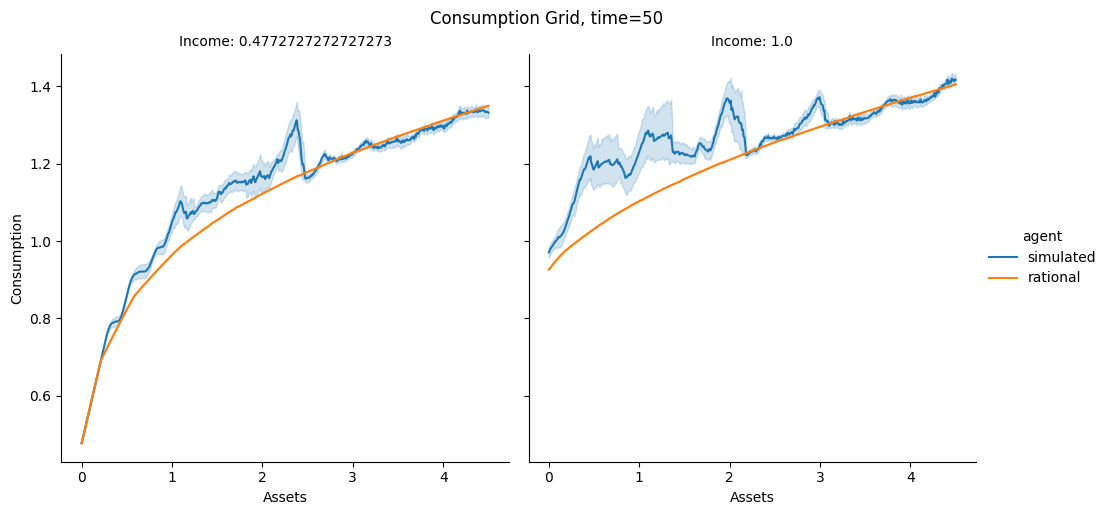}
  \caption[Consumption After 50 Periods Learning, Unsmoothed]{Consumption as a function of assets and income after 50
  periods of learning. No polynomial smoothing is used. The blue line is the reinforcement learner, the orange line is the rational benchmark. 95th percentiles are plotted. }
  \label{fig:cons-unsmooth-50}
\end{figure}

\begin{figure}[htbp!]
  \includegraphics[width=1\linewidth]{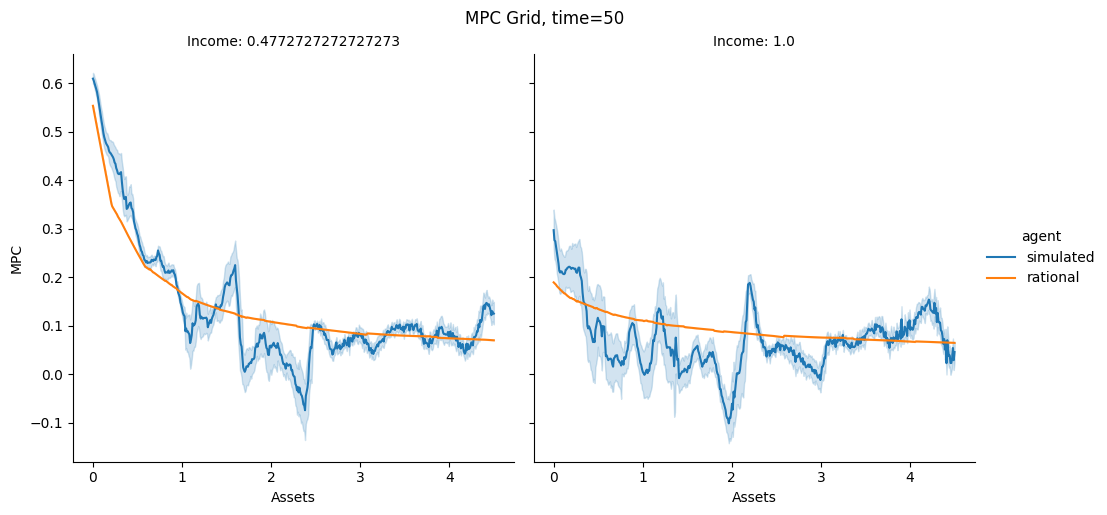}
  \caption[MPC After 50 Periods Learning, Unsmoothed]{MPC as a function of assets and income after 50 periods of
    learning. No polynomial smoothing is used. The blue line shows the reinforcement learner,
  the orange line shows the rational benchmark. 95th percentile intervals are plotted.}
  \label{fig:mpc-unsmooth-50}
\end{figure}

Consumption and MPCs without first using polynomial smoothing on the
expected value function are given below in
\cref{fig:cons-unsmooth-50} and \cref{fig:mpc-unsmooth-50}. They are
broadly similar, except without guarantees of monotonicity.
\subsection{Further Details on Experiment with Extreme Shocks}
\begin{figure}
  \centering
  \includegraphics[width=1\linewidth]{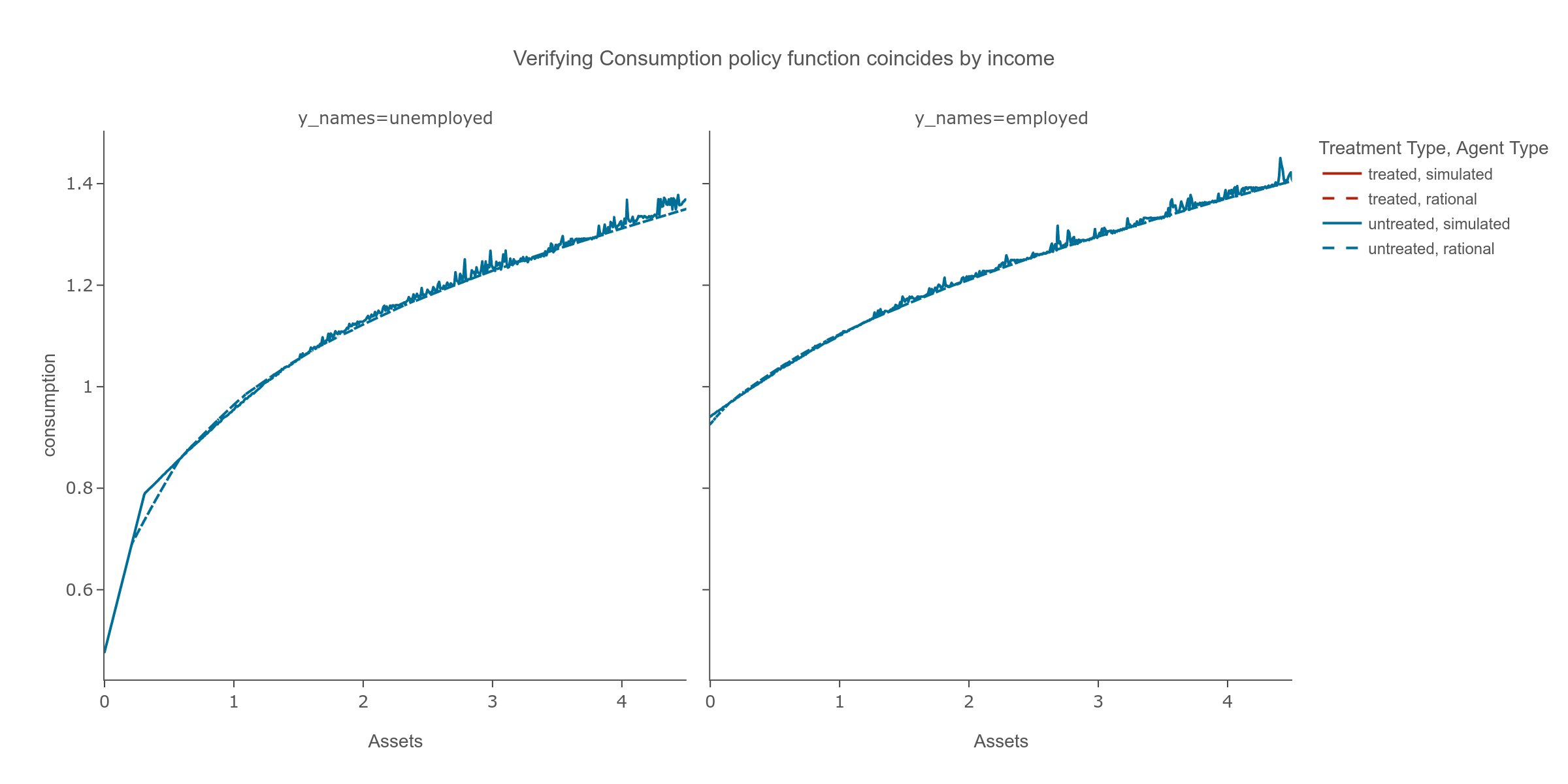}
  \caption[Consumption in Repeated Employment and Unemployment Scenario, 5 Periods]{Consumption policies at \(t = 0\) for the experiment with repeated employment with one agent receiving a series of unemployment realizations and the
    other receiving a series of employment realizations. The blue line is
    the agent with unemployment experiences, while the red line is the  agent
    with employment experiences. The dotted gray line is the rational
    agent's policy. The figure on the left represents the agent's
    consumption policy as a function of their assets (x-axis) and income
    while employed \(y_e\), while on the right it represents the agent's consumption policy as a function of their assets and income while
  unemployed \(y_u\).}
  \label{fig:initial-cons-policy}
\end{figure}

See \cref{fig:initial-cons-policy} for the initial consumption policy fit for the experiment with extreme shocks, which shows that agents began with a tight-fit to the rational policy and with identical consumption policies.
\subsection{Relaxing Markovian assumptions}\label{app:relaxing-mdp}
Three approaches stand out:
\begin{enumerate}[label=\arabic*), nosep]
\item directly modifying a Markovian algorithm, via a latent factor, conditional on which the problem again becomes Markovian
\item a separate learned representation of the environment;
\item  a model-selection based approach
\end{enumerate}

\paragraph{Latent Factor Representation}
Perhaps the most standard way is by taking all observations and capturing them in a low dimensional latent variable representing an agent's history $h_t$. This is commonly used in a non-Markovian setting called a Partially Observable Markov Decision Process (POMDP), where it is assumed that while observables may evolve in a non-Markovian fashion, there is some underlying latent hidden Markov state under which the model again becomes Markovian. To capture history dependence or limited observation, the agent is modified to use a recurrent neural network, and when updating their estimate for the continuation value, would use backpropagation through time to update their estimate of the sequence of past latent states simultaneously. More recent work has leveraged selective ``attention''-based mechanisms, or diffusion-based mechanisms. See \textcites{Tennenholtz2023, Du2024} for examples.  \textcite{Leong2017} further show that humans neurologically use attention to selectively weight values while learning about their environment via reinforcement learning, which is a natural possible future direction of research. This is not attempted  in this paper due to the additional layers of complexity and obscuring of the mechanism that would occur, and given this is meant to be an initial application of these approaches. This shares spiritual similarities with filtering, including the Kalman filter, and other estimation procedures for Hidden Markov Models.

\paragraph{Model of Environment}
A second approach would be to model the agent's environment in a flexible manner using a low-dimensional representation for observations, updating the model over time. This is what \textcite{Hafner2025} does recently, one of the first at-scale model-based RL procedures, using encoded latent factor representations to capture a ``world model'' generating observations. Alternatively, other approaches can be used to perform some kind of model selection procedure on a set of states. Two recent examples exploring this in a setting similar to my paper are \textcites{Levine2024, Lamb2023}. These are  more sophisticated in that they not only find a relevant subset of states, but also identify over time which states are controllable versus exogenous.

\paragraph{Non-Markovian}
These methods can also be applied to relax Markov settings to hidden Markov models, conditionally Markovian settings in continuous time, or fully non-Markovian settings. See \textcites{Whitehead1995,Kaelbling1998, Sutton1999, Qin2023} for further details.
\subsection{Changing Income Probabilities Experiment}\label{app:pessimism}
This appendix provides full details for the pessimism comparison summarized in \cref{tab:rl-vs-pessimism} in the main text.

\begin{figure}[htbp!]
\centering
\includegraphics[width=1\linewidth]{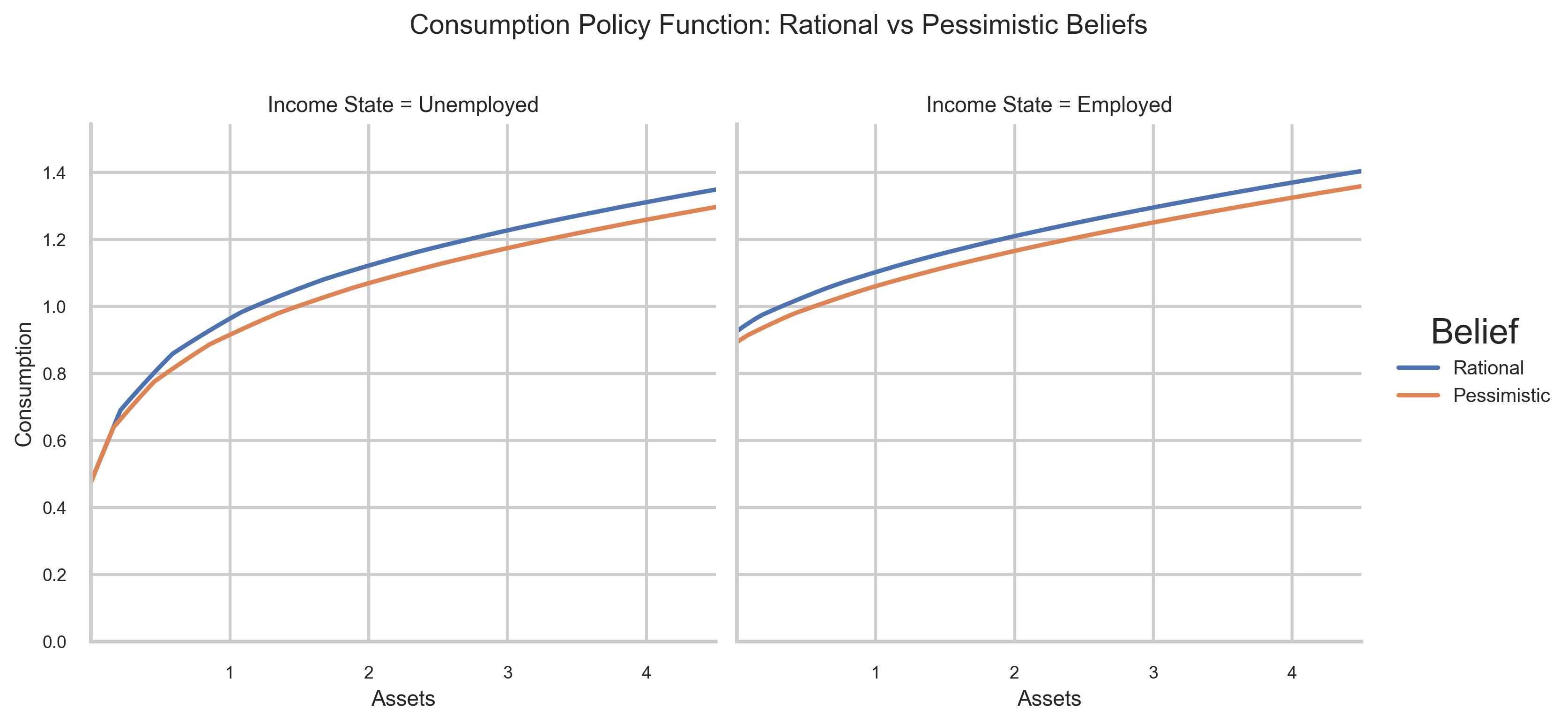}
\caption[Consumption in Response to Pessimism]{Orange line is rational agent solution under 1.5x pessimistic income transition probabilities, blue line is rational agent solution under original income transition probabilities.  The figure on the left represents the agent's consumption policy as a function of their assets (x-axis) and income while unemployed \(y_u\), while on the right it represents the agent's consumption policy as a function of their assets and income while unemployed \(y_e\).}
\label{fig:cons-pessimism}
\end{figure}

\begin{figure}[htbp!]
\centering
\includegraphics[width=1\linewidth]{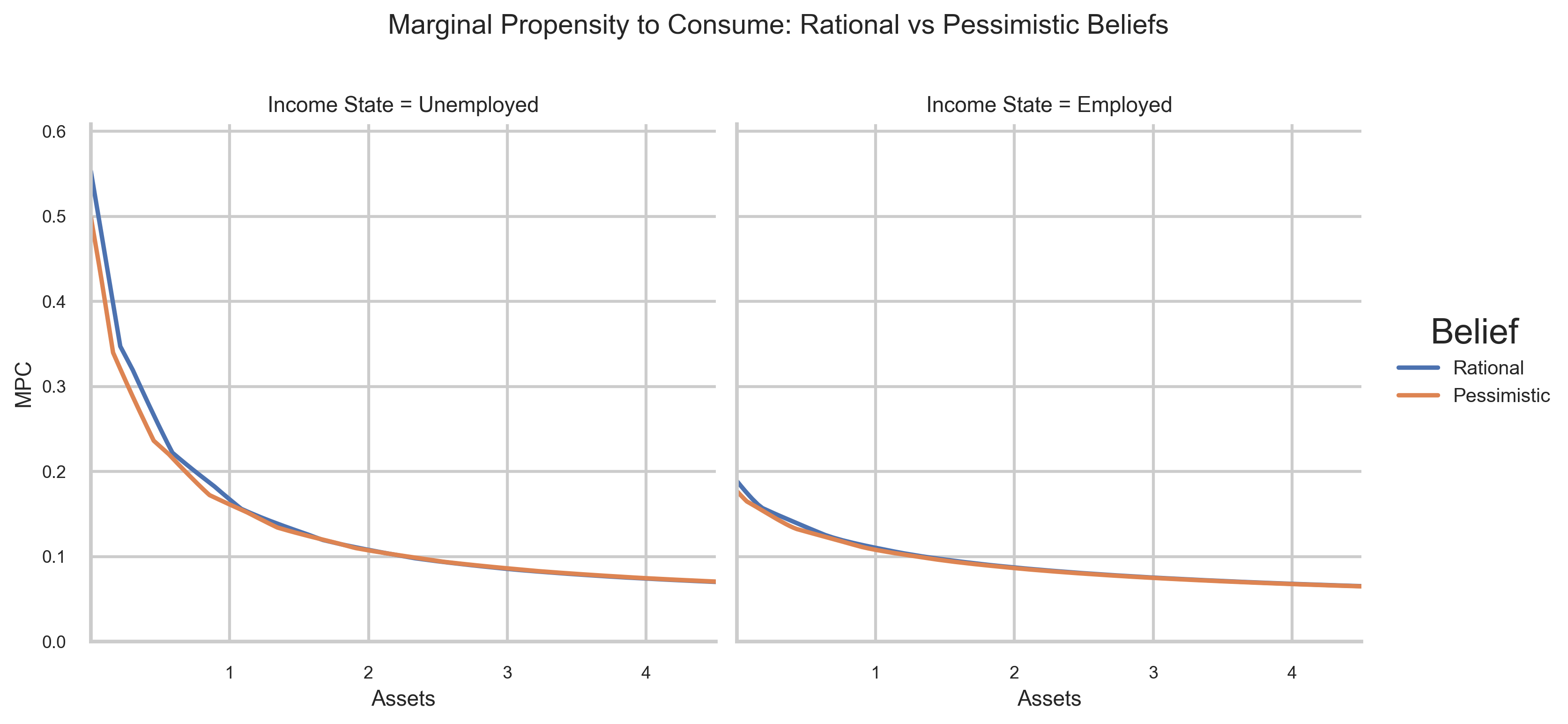}
\caption[Consumption in Response to Pessimism]{Orange line is rational agent marginal propensity to consume under 1.5x pessimistic income transition probabilities, blue line is rational agent solution under original income transition probabilities.  The figure on the left represents the agent's consumption policy as a function of their assets (x-axis) and income while unemployed \(y_u\), while on the right it represents the agent's consumption policy as a function of their assets and income while unemployed \(y_e\).}
\label{fig:mpc-pessimism}
\end{figure}

I study what happens under the current parameterization in response to an increase in perceived transitions to unemployment, but without approximating the conditional expected value function directly. I use the existing parameterization \cref{tab:all-params}, but solely run the rational model under two specifications, one with the original income transition probabilities, and one where the $UU$ and $EU$ probabilities are increased by 1.5 times their original value, and then the probabilities are re-normalized to sum to 1. Crucially, for each run, I use the perceived income distribution, but then solve the policy via iterating forward to a fixed point under this perceived income distribution, as in \textcite{Malmendier2024}. The results are shown in \cref{fig:cons-pessimism} and \cref{fig:mpc-pessimism}. This shows a tilt down globally in the consumption policy for the rational agent and a shift down in the marginal propensity to consume, as the agent is more pessimistic about their future income prospects. This is consistent with the intuition that pessimism leads to lower consumption, as the agent is more likely to be unemployed in the future and so saves more for that eventuality. Because the agent exhibits prudence in their utility function, a lower consumption policy everywhere in turn leads to a lower local slope of the consumption policy and lower MPCs per asset level.\footnote{Even without prudence, this could occur strictly  due to the presence of the borrowing constraint. However, the author knows of no theoretical result that guarantees this at the time of writing.}

\subsection{Long Run and Convergence}
\begin{figure}
  \includegraphics[width=1\linewidth]
  {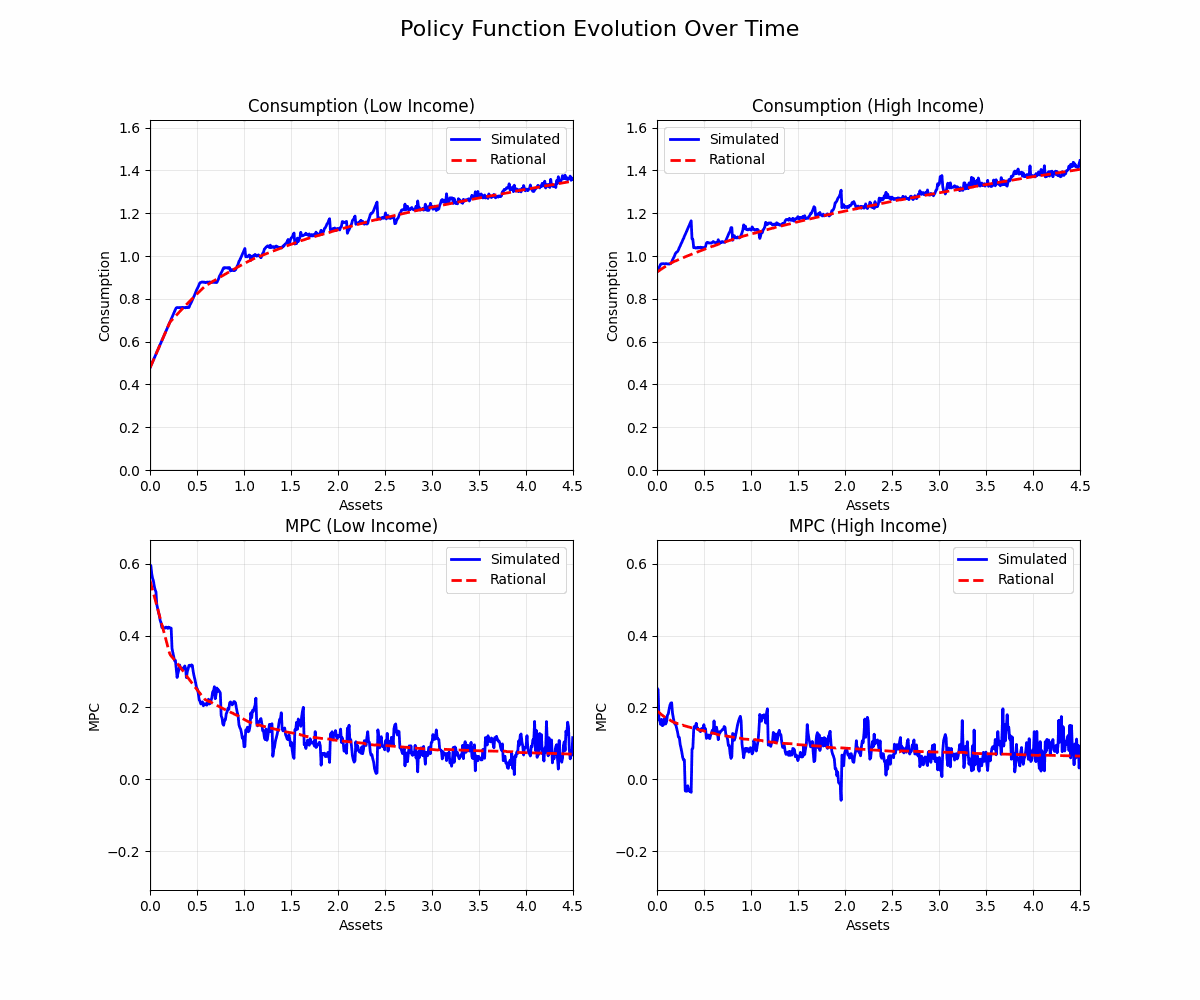}
  \caption[Initial Consumption Policy, MPC Fit]{Initial consumption policy fit at \(t = 0\) quarters for a single seed, no smoothing.}
  \label{fig:pol-0}
\end{figure}

\begin{figure}
  \includegraphics[width=1\linewidth]
  {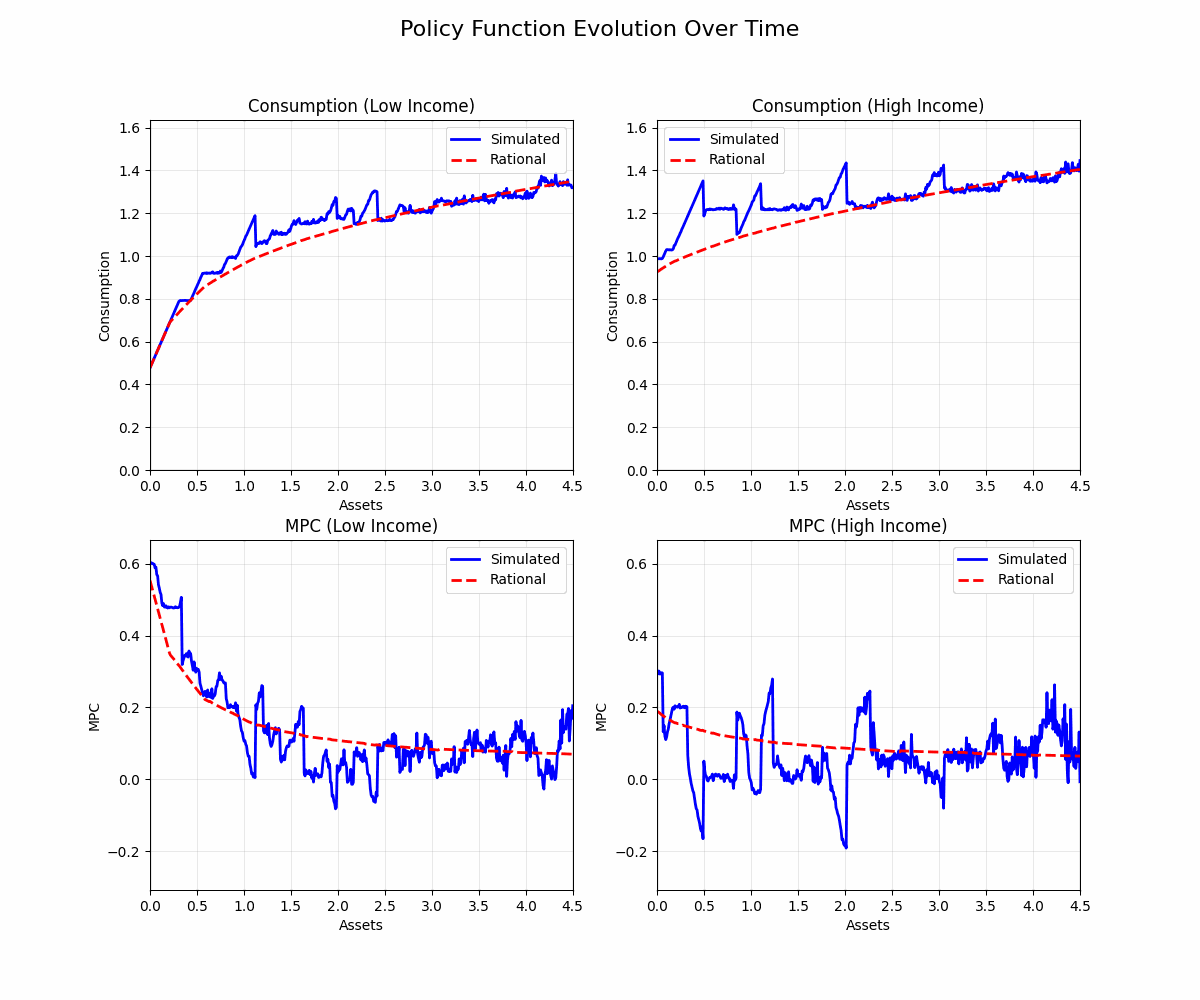}
  \caption[Consumption Policy, MPC Fit at \(t = 10\)]{Consumption policy and MPC fit against rational at \(t = 10\) quarters for a single seed, no smoothing, systematic tilting of policy function away from rational, as well as local fluctuations without smoothing.}
  \label{fig:pol-1}
\end{figure}

\begin{figure}
  \includegraphics[width=1\linewidth]
  {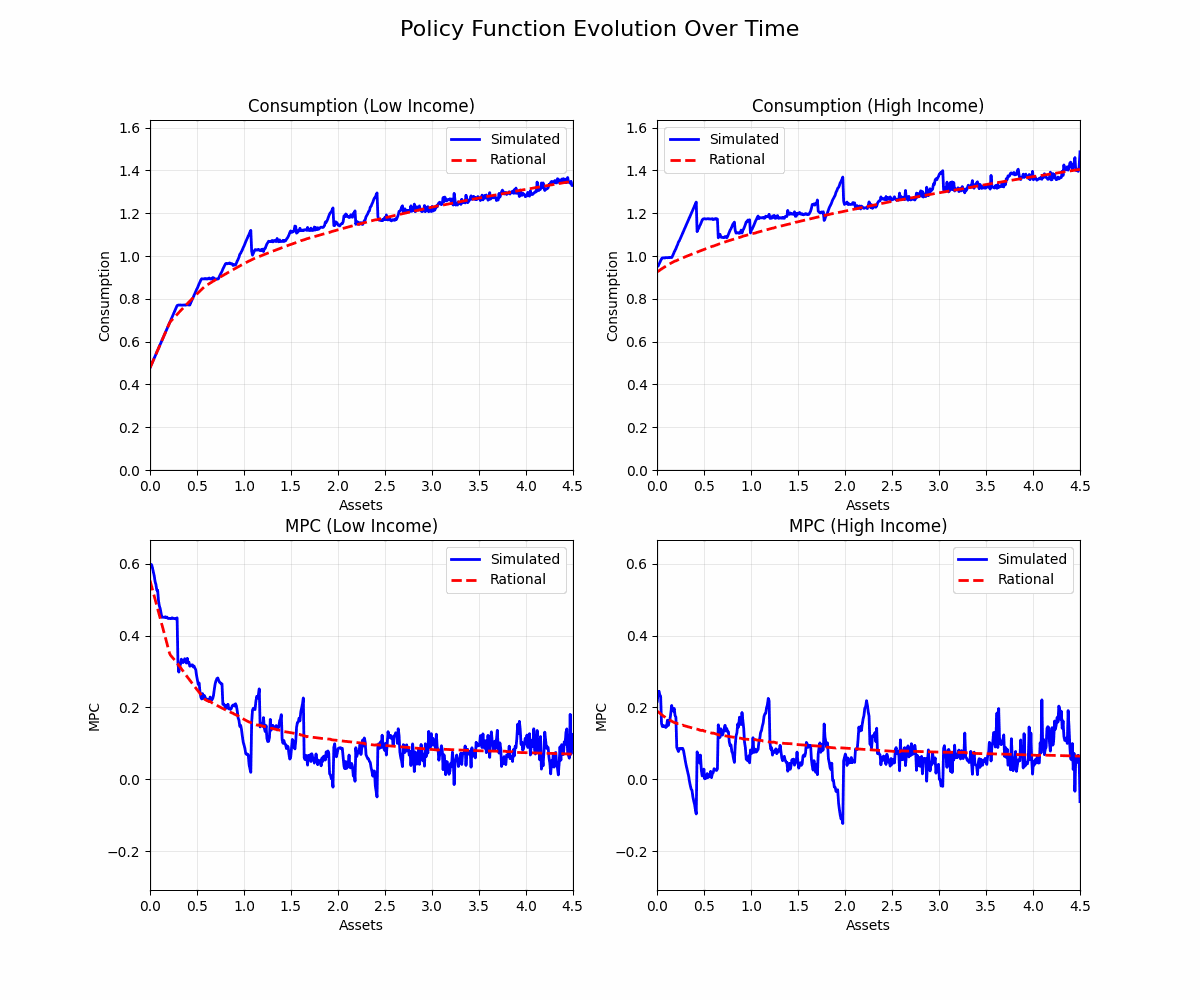}
  \caption[Consumption Policy, MPC Fit at \(t = 240\)]{Consumption policy and MPC fit against rational at \(t = 240\) quarters for a single seed, no smoothing. Local fluctuations remain elevated but systematic bias in policy has mostly vanished.}
  \label{fig:pol-14}
\end{figure}

Policies appear to exhibit convergence in the long-run. In figures \cref{fig:pol-1}, the policy is examined at time 10, in \cref{fig:pol-14} at period 240. Both are compared to time 0 initial fit in \cref{fig:pol-0}. Policies first drift away from rational solution before converging back.

\printbibliography
\end{document}